\documentclass[prd,aps,twocolumn,floats,balancelastpage,superscriptaddress,showkeys,nofootinbib,showpacs,preprintnumbers,floatfix]{revtex4-1}
\usepackage[english]{babel}
\usepackage[utf8]{inputenc}
\usepackage{graphicx, epsfig, bm}
\usepackage{xcolor}
\usepackage{graphicx,amsmath,amsfonts, soul, amssymb,color,float,wasysym,wrapfig,xspace,changepage,lineno,todonotes}
\definecolor{colorLink}{rgb}{0.7,0,0}
\definecolor{colorCite}{rgb}{0,.7,0}
\definecolor{colorURL}{rgb}{0,0,0.7}
\usepackage[colorlinks=true
,urlcolor=blue
,anchorcolor=blue
,citecolor=blue
,filecolor=blue
,linkcolor=red
,menucolor=blue
,linktocpage=true
,pdfproducer=medialab
,pdfa=true
]{hyperref}
\usepackage{setspace}
\usepackage{multirow}
\usepackage{placeins}
\usepackage{array}

\usepackage{academicons}

\newcommand{\orcidlink}[1]
{\begingroup
  \hypersetup{hidelinks}\href{https://orcid.org/#1}{\includegraphics[width=10pt]{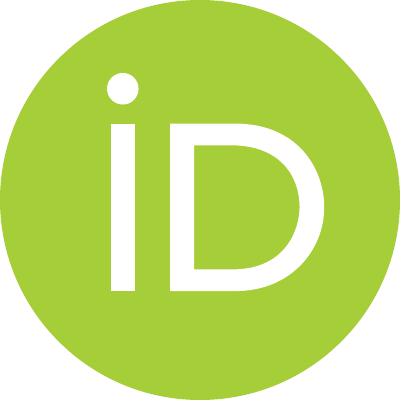}
} \endgroup}

\newcommand{\DM}{{\scriptscriptstyle \text{DM}}}
\newcommand{\NFW}{{\scriptscriptstyle \text{NFW}}}
\newcommand{\Einasto}{{\scriptscriptstyle \text{Einasto}}}
\newcommand{\cNFW}{{\scriptscriptstyle \text{cNFW}}}
\newcommand{\mW}{m_{\scriptscriptstyle W}}
\def\mnras{Mon.~Not.~Roy.~Astron.~Soc.} 

\begin{document}

\preprint{CERN-TH-2022-156; IRFU-22-02}

\title{Toward the ultimate reach of current imaging atmospheric Cherenkov telescopes \\ and their sensitivity to TeV dark matter}

\date{\today}

\author{Alessandro Montanari \orcidlink{0000-0002-3620-0173}}
\affiliation{\!\!\mbox{ \footnotesize IRFU, CEA, D\'epartement de Physique des Particules, Universit{\'{e}} Paris-Saclay, F-91191 Gif-sur-Yvette, France}\,\,\,\vspace{0.7ex}}
\author{Emmanuel Moulin \orcidlink{0000-0003-4007-0145}}
\affiliation{\!\!\mbox{ \footnotesize IRFU, CEA, D\'epartement de Physique des Particules, Universit{\'{e}} Paris-Saclay, F-91191 Gif-sur-Yvette, France}\,\,\,\vspace{0.7ex}}
\author{Nicholas L. Rodd \orcidlink{0000-0001-7559-9597}}
\affiliation{\!\!\mbox{ \footnotesize Theoretical Physics Department, CERN, 1 Esplanade des Particules, CH-1211 Geneva 23, Switzerland}\,\,\,\vspace{0.7ex}}

\begin{abstract}
Indirect detection opens a unique window for probing thermal dark matter (DM): the same annihilation process that determined the relic abundance in the early Universe drives the present day astrophysical signal. While TeV-scale particles weakly coupled to the Standard Model face undoubted challenges from decades of null searches, the scenario remains compelling, and simple realizations such as Higgsino DM remain largely unexplored. The fate of such scenarios could be determined by gamma-ray observations of the centre of the Milky Way with Imaging Atmospheric Cherenkov Telescopes (IACTs). We consider the ultimate sensitivity of current IACTs to a broad range of TeV-scale DM candidates---including specific ones such as the Wino, Higgsino, and Quintuplet. To do so, we use realistic mock H.E.S.S.-like observations of the inner Milky Way halo, and provide a careful assessment of the impact of recent Milky Way mass modeling, instrumental and astrophysical background uncertainties in the Galactic Center region, and the theoretical uncertainty on the predicted signal. We find that the dominant systematic for IACT searches in the inner Galaxy is the unknown distribution of DM in that region, however, beyond this the searches are currently statistically dominated indicating a continued benefit from more observations. For two-body final states at $1~{\rm TeV}$, we find a H.E.S.S.-like observatory is sensitive to $\langle \sigma v \rangle \sim 3 \times 10^{-26}-4 \times 10^{-25}~{\rm cm}^3{\rm s}^{-1}$, except for neutrino final states, although we find results competitive with ANTARES. In addition, the thermal masses for the Wino and Quintuplet can be probed; the Higgsino continues to be out of reach by at least a factor of a few. Our conclusions are also directly relevant to the next generation Cherenkov Telescope Array, which remains well positioned to be the discovery instrument for thermal DM.
\end{abstract}

\keywords{dark matter, gamma rays, Galactic center, Galactic halo}
\pacs{
95.35.+d, 95.85.Pw, 98.35.Jk, 98.35.Gi
}
\maketitle

\section{Introduction}
\label{sec:introduction} 

An overwhelming body of astrophysical and cosmological measurements show that close to 85\% of the total matter content of the Universe consists of non-baryonic dark matter (DM).
However, its nature remains unknown.
Although the space of DM models remains vast, an electroweak scale thermal relic or WIMP has long been identified as the archetypal DM candidate for its economy and elegance~\cite{Jungman:1995df}.
Null searches carried out in direct detection experiments and at colliders (see e.g. Refs.~\cite{Schumann:2019eaa,Kahlhoefer:2017dnp}) have excluded many realizations of the scenario, particularly for masses near the electroweak scale ($m_{\DM} \sim 100~{\rm GeV}$).
Although constrained, GeV-scale WIMPs remain far from excluded~\cite{Leane:2018kjk}.
As one moves to TeV masses, long predicted DM candidates, including the Higgsino, remain not only untested, but out of the reach of near-term terrestrial probes (for a recent overview, see, e.g., Refs.~\cite{Bottaro:2021snn,Bottaro:2022one}).
TeV WIMPs can, however, be shining brightly in very-high-energy (VHE, E $\gtrsim$ 100 GeV) gamma rays, opening the potential for Imaging Atmospheric Cherenkov Telescopes (IACTs) to be DM discovery instruments.

Constraints on heavy WIMPs arise from a variety of sources, including nearby dwarf galaxy observations from Fermi-LAT~\cite{Fermi-LAT:2015att,PhysRevD.104.083026} and IACTs~\cite{HESS:2020zwn,HESS:2021zzm,MAGIC:2021mog}, Galactic Center (GC) observations by Fermi-LAT~\cite{Fermi-LAT:2017opo}, HAWC~\cite{HAWC:2017udy}, and IACTs~\cite{Abdallah:2016ygi,Abdallah:2018qtu}, as well as from CMB measurements~\cite{Planck:2018vyg}.
At present, some of the strongest constraints derived on the annihilation rate of TeV WIMPs are obtained from 546 hours of H.E.S.S. full-array observations of the GC~\cite{HESS:2022ygk}.
Beyond any possible signal of DM, the GC at TeV energies is a rich and challenging environment, often associated with extreme astrophysical phenomena.
Faint diffuse emission has been detected, including VHE diffuse emission likely attributed to PeV hadrons accelerated in the vicinity of the supermassive black hole Sgr A*, referred to as a Pevatron~\cite{Abramowski:2016mir}, and the base of the Fermi Bubbles~\cite{Su:2010qj,Acero:2016qlg,Fermi-LAT:2017opo} for which the origin is still unknown.
In addition, the Galactic Centre excess (GCE) detected in Fermi-LAT data~\cite{Goodenough:2009gk,Hooper:2010mq,Fermi-LAT:2015sau} could be attributed to a putative millisecond pulsar population in the inner Galaxy~\cite{Lee:2015fea,Bartels:2015aea}, a population that could induce VHE gamma-ray emissions via electron inverse Compton scattering off ambient radiation fields~\cite{Song:2021zrs}.
While the GC is undoubtedly complex, it is also the region where we expect the brightest signal of DM annihilation, and is therefore the focus of many IACT DM searches, for example Refs.~\cite{Abdalla:2016olq,Weinstein:2016vyv,Lacroix:2016qag,Abdallah:2018qtu,MAGIC:2021frm,HESS:2022ygk} including the central VHE source HESS J1745-290~\cite{Belikov:2012ty,Cembranos:2013fya} as well as detection prospects with CTA~\cite{Lefranc:2015pza,Belikov:2016fwv,Acharyya:2020sbj,Rinchiuso:2020skh}, and of the present work.

The particular focus of this work is look towards establishing the ultimate sensitivity of the current generation of IACTs to TeV DM annihilation, with a particular focus on quantifying the possible limiting systematics.
We will do so by using mock data from H.E.S.S.-like observations of the GC with the five-telescope array, and by deriving sensitivities with various limiting systematics included.
Our analysis will combine the following three elements: {\it (i)} the most precise calculations available for the gamma-ray annihilation yield for a wide range of DM masses and models in order to estimate the theoretical uncertainty on the signal; {\it (ii)} updated models for the Milky Way mass profiles, specifically those from Ref.~\cite{Cautun:2019eaf} which drew on Gaia measurements; {\it (iii)} realistic modeling of the TeV astrophysical backgrounds in the GC together with appropriated uncertainties.
We will apply this approach to compute the IACT sensitivity in the $0.5$ to $100~{\rm TeV}$ mass range for both model-independent two-body final states searches (including the two-neutrino channels), as well as specific analyses for the Wino, Higgsino, and Quintuplet models.
(In this context final state refers to the particles emerging from the hard annihilation process.
The eventual states that result once showering and hadronization take place which we are interested in are, of course, photons.)
In all cases, we will use a log-likelihood-ratio test statistic analysis that incorporates both spatial and energy information.

Let us already summarize the key findings that will be demonstrated by our analysis.
\begin{itemize}
    \item The dominant systematic uncertainty on DM searches in the GC remains the modeling of the DM distribution in that region.
    \item Beyond this, however, the analysis becomes statistics dominated, implying that continued data collection with existing IACTs remains important.
    \item IACTs are broadly sensitive to TeV DM candidates, but provide a unique probe of the thermal Wino and Quintuplet, and we demonstrate that a unique feature of the latter is a rapid variation in the shape of the spectrum as a function of mass.
    \item The thermal Higgsino remains out-of-reach for existing instruments given the assumptions adopted in this work.
    \item Galactic diffuse emission is already a relevant background for H.E.S.S.-like searches.
\end{itemize}
Importantly, we note that while the above conclusions and our focus relate to existing IACTs, these lessons will be directly applicable to the upcoming Cherenkov Telescope Array, which in addition to its broad DM reach, will be the first instrument in reach of the thermal Higgsino~\cite{Hryczuk:2019nql,Rinchiuso:2020skh}.
For the deep observational program that will be carried out in the GC by the Cherenkov Telescope Array (CTA) to probe thermal WIMPs~\cite{Moulin:2019oyc,CTA:2020qlo}, an optimal control of the observational and instrumental systematic uncertainties will be required to make the most of the massive expected dataset, exactly the issues we consider in this work.
However, we emphasize that in the next few years H.E.S.S. will likely have amassed over 1,000 hours of observations near the GC, suggesting that CTA should look to target even larger observations.

The paper is organized as follows.
Section~\ref{sec:sources} describes the various sources of gamma-ray flux in the GC, including both the putative of WIMP annihilation, but also astrophysical emission mechanisms that will form the backgrounds to DM searches.
In Sec.~\ref{sec:analysis}, we outline the analysis procedure we will use to search for DM and determine the systematic uncertainties associated with those analyses, leaving the results to Sec.~\ref{sec:results}.
Finally, we devote Sec.~\ref{sec:conclusions} to our conclusions.

\section{Sources of TeV Photons in the Inner Galaxy}
\label{sec:sources}

We begin our discussion with the various contributions expected for TeV gamma-rays in the inner region of our Galaxy.
We will explore the various aspects associated with the possible DM flux, including the spectrum of photons produced by DM annihilation, and also the uncertainties associated with the DM distribution in this region.
Additionally, we will describe the various backgrounds expected from the GC.

\subsection{Gamma rays from dark-matter annihilation}
\label{subsec:SigSpec_thmodels}

If the DM in the Milky Way halo is made of self-conjugate WIMPs undergoing self-annihilation, then there will be a potentially detectable flux of gamma rays incident on the Earth.
Searching for this flux is generally characterized as the indirect detection of DM.
If we observe a solid angle of the sky $\Delta\Omega$, then the energy-differential flux is given by
\begin{equation}\begin{aligned}
\label{eq:dmflux}
\frac{d \Phi}{d E} &=
\frac{\langle \sigma v \rangle}{8\pi m_{\DM}^2}\sum_f  {\rm BR}_f \frac{d N_f}{d E} \, J(\Delta\Omega), \\
J(\Delta\Omega) &= \int_{\Delta\Omega} d \Omega \int_{\rm los}  d s\, \rho_\DM^2(s[r,\theta]).
\end{aligned}\end{equation}
The flux as determined by the above expression depends on both fundamental details of the particle physics, which controls the annihilation, and further the astrophysical distribution of the DM itself.
We will detail the various ingredients below.

Firstly, the observed flux depends on the distribution of DM in the Milky Way through the $J$-factor, $J(\Delta\Omega)$, which corresponds to the integral of the square of the DM density $\rho_\DM$ over the line of sight (los) $s$ and solid angle $\Delta\Omega$.
As we have written the $J$-factor, the DM density $\rho_\DM$ is assumed to be distributed spherically symmetrically around the GC, and therefore $s$ depends only on the radial distance from the GC, $r$, and the angle of the los from the GC, $\theta$.
To be explicit, $r^2 = s^2 +r_{\odot}^2-2\,r_{\odot}\,s\, \cos\theta$, where we take the distance between our observations and the GC to be $r_\odot = 8.127~{\rm  kpc}$~\cite{GRAVITY:2018ofz}.
In order to compute $J(\Delta \Omega)$ a model for the DM density within the Milky Way, $\rho_\DM(r)$, is required, and we will discuss the forms we adopt and their associated uncertainties in Sec.~\ref{subsec:DM_distr}.

The DM flux in Eq.~\eqref{eq:dmflux} further depends on detailed aspects of the particle physics that describes the annihilation.
The ingredients include the velocity-weighted annihilation cross-section, $\langle \sigma v \rangle$, the DM mass, $m_{\DM}$, and the energy-differential yield of gamma rays per annihilation (or spectrum), $d N_f/d E$, where $f$ denotes the annihilation channel, which are then weighted by the branching ratio to each channel, ${\rm BR}_f$.
A complete first-principles description of these factors requires a full DM model, describing all the interactions between DM and the Standard Model.
Armed with such a complete theory, one could in principle compute to the desired accuracy the annihilation rate, branching fractions to different final states, and the associated spectra of each, all as a function of the underlying parameters of the ultraviolet theory.
This is exactly the approach we will pursue for several canonical WIMP candidates in the next subsection.
However, one can also adopt a more model-independent mindset where searches are performed for specific final states, i.e. a specific $f$, and then repeated across a representative set of final states.
A detection or limit on the flux associated with a specific channel can then be interpreted as a constraint on $\langle \sigma v \rangle \times {\rm BR}_f$, which then can be interpreted in terms of specific models.
As we review in the next subsection, the thermal relic prediction of $\langle \sigma v \rangle \sim 10^{-26}~{\rm cm}^3{\rm s}^{-1}$ represents an important model-independent target~\cite{Steigman2012}.
The key ingredient in the model-independent approach is then a representative set of final states $f$, and a precise calculation of the spectra for each, $dN_f/dE$.
For the choice of $f$, we follow the standard approach of considering a broad range of two-body final states, such as annihilation to $W^+ W^-$.
The justification is that unless the two-body final states are inaccessible, they should dominate as processes with more final state particles will be phase-space suppressed.
(Note that while in practice moving beyond two-body final states considerably opens the number of possibilities, certain classes of scenarios, such as cascades in the dark sector, are straightforward to incorporate -- see e.g. Refs.~\cite{Mardon:2009rc,Elor:2015tva,Elor:2015bho,Baldes:2020hwx} -- although we will not do so here.)
The specific two-body final states we will consider are $W^+ W^-$, $ZZ$, $HH$, $b \bar{b}$, $t \bar{t}$, and $\ell^+ \ell^-$ as well as $\nu_{\ell} \bar{\nu}_{\ell}$ for $\ell=e,\mu,\tau$.

Whilst the possible mass range for DM is known to be vast, in this work we will focus our attention on masses between 500 GeV and 100 TeV.
The lower bound is prompted by the strong constraints imposed by non-observation of DM annihilation signals from, for example, dwarf spheroidal galaxies by Fermi-LAT~\cite{Ackermann:2015zua}.
The upper end of our mass range is motivated by the unitarity bound~\cite{Griest:1989wd}, although only weakly---unitarity applied to point-like particle annihilation of self-conjugate DM in a conventional thermal cosmology requires $m_{\DM} < 194~{\rm TeV}$~\cite{Smirnov:2019ngs,Tak:2022vkb}.
Of course, the unitarity bound is straightforward to evade, even without modifying the underlying cosmology. 
If the underlying particle physics is modified to include compositeness or bound state formation, much heavier DM can be accommodated (for a recent discussion, see for example, Refs.~\cite{Bottaro:2021snn,Carney:2022gse,Tak:2022vkb}).
Especially as the next-generation IACT CTA comes online, searches for DM at even heavier masses will be particularly worthwhile~\cite{Moulin:2019oyc,Tak:2022vkb}.
Throughout this work we will show no preference for any particular $m_{\DM}$ within the range we consider, although in specific DM models a theoretical preference for certain masses can emerge, and we give examples in Sec.~\ref{subsec:TeVmodels}.

\begin{figure}[!t]
\includegraphics[width=0.47\textwidth]{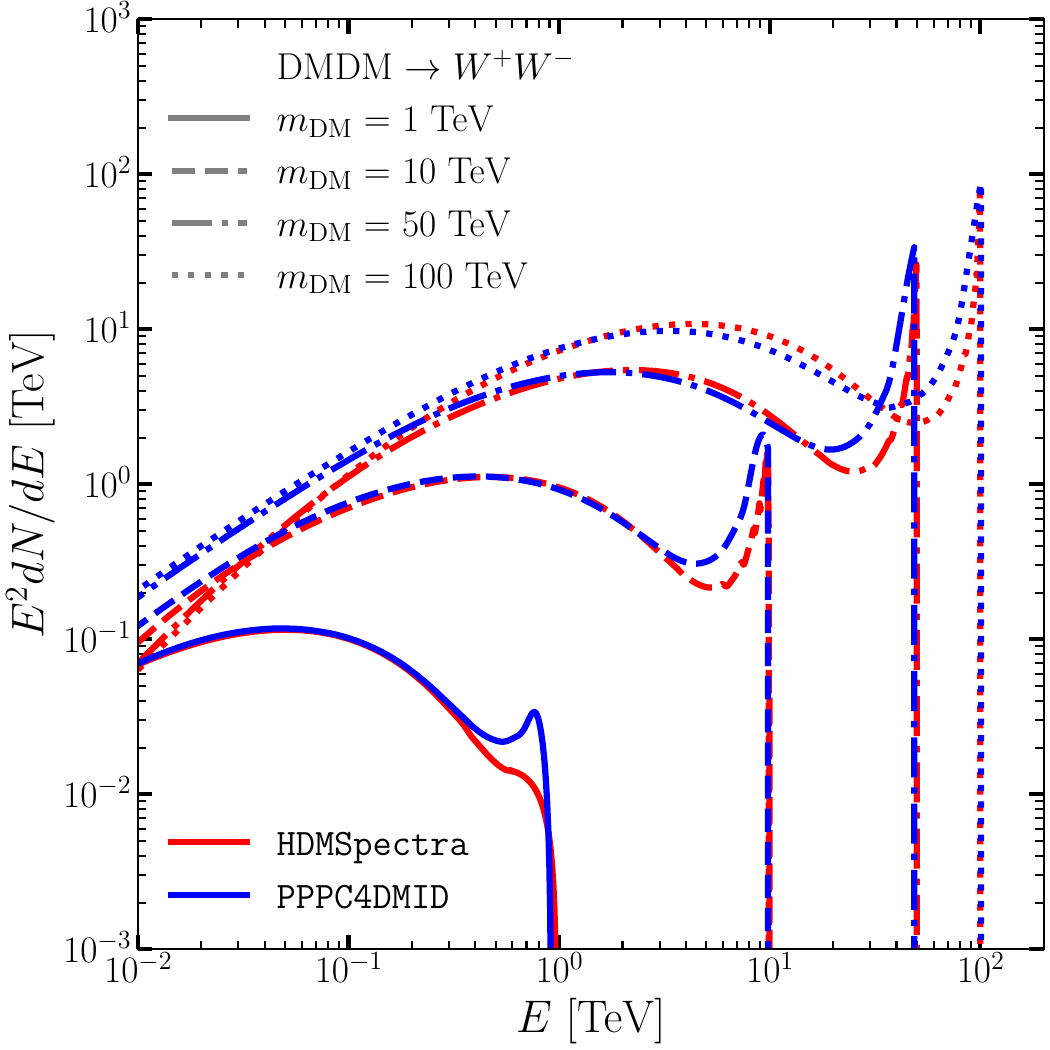}
\caption{The spectrum of gamma-rays emerging per annihilation for DM which annihilates solely to a $W^+ W^-$ final state.
Results are shown for four masses: 1 (solid lines), 10 (dashed lines), 50 TeV (dashed-dotted lines), and 100 TeV (dotted lines), and in each case we show the spectra as provided by \texttt{PPPC4DMID}~\cite{Cirelli:2010xx} (blue lines) and \texttt{HDMSpectra}~\cite{Bauer:2020jay} (red lines).
We use the differences between these approaches to estimate the theoretical uncertainties in our model-independent approach.}
\label{fig:GammaYield}
\vspace{-0.5cm}
\end{figure}

The final ingredient required to complete our description of Eq.~\eqref{eq:dmflux} is the differential energy spectra $dN_f/dE$.
Here we focus on the model-independent two-body final states, again leaving the results for specific models to the following subsection.
The physics involved in the computation of these spectra is rich.
For instance, consider determining the photon spectrum that results from DM annihilation to $b \bar{b}$.
This requires an accounting for the QCD showering describing, for instance, the emission of gluons from the quarks and the resulting hadronization of the system, as a significant number of the photons in the spectrum will arise from the decay of neutral pions produced in this process.
For $m_{\DM}$ well above the electroweak scale, the emission of the electroweak bosons can also become relevant for determining the ultimate spectrum~\cite{Ciafaloni:2010ti}, indeed at these masses even two-body neutrino final states do not simply produce a neutrino line, but can further produce a considerable flux of photons, which is why we will consider them.
In particular, this allows IACTs to probe final states that are traditionally the focus of neutrino telescopes like IceCube and ANTARES, and indeed we will show results comparing the two approaches.
(Broadly, these effects imply that searches for heavier DM is inherently multimessenger, see e.g. Refs.~\cite{Murase:2012xs,Ahlers:2015moa,Cohen:2016uyg,Cembranos:2019noa,Gammaldi:2019mel,Carney:2022gse}.)
Accordingly, results that incorporate all of the above physics are required for an accurate determination of the spectrum.
By default we adopt the recent \texttt{HDMSpectra} computation from Ref.~\cite{Bauer:2020jay}, however, in order to estimate the systematic dependency of our conclusions on this choice, we will compare our findings to results obtained using the alternative approach from \texttt{PPPC4DMID}~\cite{Cirelli:2010xx}.
Both of these approaches rely on \texttt{Pythia}~\cite{Sjostrand:2006za,Sjostrand:2007gs,Sjostrand:2014zea} for describing the evolution of the system below the electroweak scale, however the evolution at higher energies is treated differently between the two.
For instance, \texttt{PPPC4DMID} includes finite electroweak mass effects, which are relevant for $m_{\DM}$ near the electroweak scale, whereas \texttt{HDMSpectra} allows for multiple emissions of electroweak bosons that are of growing importance as $m_{\DM}$ becomes heavier and heavier.
Figure~\ref{fig:GammaYield} provides a comparison of the two approaches for the $W^+ W^-$ final state, where the different electroweak effects are relevant, and across the range of masses we consider.
As the figure shows, there can be clear differences in the hard photon spectra across the full mass range, and we will use the difference as a proxy for the theoretical uncertainty that exists at present on the spectra.
This estimate will not include uncertainties associated with ingredients common to both approaches, such as the QCD uncertainties associated with the use of \texttt{Pythia}.
These have, however, previously been estimated in Ref.~\cite{Amoroso:2018qga}, where it was seen those effects could result in systematic errors varying from a few percent to as large as 50\%.

\subsection{Canonical TeV WIMP candidates: the Higgsino, Wino and Quintuplet models}
\label{subsec:TeVmodels}

Why would we expect the DM of our Universe to be a weakly coupled electroweak scale particle?
After all, WIMPs are intimately associated with supersymmetry -- often emerging as the expected DM particle from SUSY models -- and the absence of any evidence for this framework at colliders may lead one to question the motivation for WIMPs themselves~\cite{ParticleDataGroup:2022pth,Giudice:2017pzm,Feng:2013pwa}.\footnote{We note there are other possible UV origins for the WIMP, including extra-dimensions, see for example Refs.~\cite{Cheng:2002ej,Servant:2002aq,Cembranos:2003mr}.
However, these scenarios are also challenged by collider measurements~\cite{ParticleDataGroup:2022pth}.}
While it is true that the full UV motivation for WIMPs is not what it was several decades ago, the scenario remains compelling.
There is an elegance to the thermal relic cosmology: DM emerges from the primordial plasma with the correct relic abundance if its velocity weighted annihilation cross-section is $\langle \sigma v \rangle \sim 10^{-26}~{\rm cm}^3{\rm s}^{-1}$~\cite{Steigman:2012nb}, exactly at a level where the late time annihilations in the GC could be detectable.
Being completely model-independent, it is then worthwhile to consider DM annihilations around this cross section to a broad range of final states, as we introduced in the previous subsection.

There further remains strong motivation for more specific realizations of the WIMP.
Putting UV questions aside, one can consider what are the most economical models of DM, in the sense of what is the minimal field content one could add to the Standard Model to explain DM.
The answer to this question is TeV scale states charged under the electroweak interaction, and includes an SU(2) doublet with unit hypercharge, as well as a $\mathbf{3}$ and $\mathbf{5}$ representation of SU(2)~\cite{Cirelli:2005uq,Cirelli:2007xd,Cirelli:2008id,Cirelli:2009uv,Cirelli:2015bda,Mahbubani:2005pt,Kearney:2016rng}.
These three states are also known as the Higgsino, Wino, and Quintuplet, and produce the correct DM abundance when embedded in a thermal relic cosmology if they have masses of $1.0 \pm 0.1$ TeV, $2.9 \pm 0.1$ TeV, and $13.6 \pm 0.8$ TeV, respectively~\cite{Cirelli:2007xd,Hisano:2006nn,Hryczuk:2010zi,Beneke:2016ync,Mitridate:2017izz,Bottaro:2021snn}.
(For a broad consideration of the detection prospects for these minimal DM candidates, and others, see Refs.~\cite{Bottaro:2021snn,Bottaro:2022one}.)
Intriguingly, the Higgsino and Wino are also the thermal DM candidates that often emerge from UV scenarios realizing supersymmetry in a manner consistent with LHC observations -- usually at the cost of a natural solution to the hierarchy problem -- such as split~\cite{ArkaniHamed:2004fb,Giudice:2004tc,ArkaniHamed:2004yi,Wells:2004di,Pierce:2004mk,Arvanitaki:2012ps,ArkaniHamed:2012gw,Fox:2014moa} and spread~\cite{Hall:2011jd,Hall:2012zp} SUSY.
While there remain various long term paths to discover the DM in these scenarios (see e.g. Refs.~\cite{Co:2021ion,Co:2022jsn}), that CTA could soon see a signal from the Higgsino~\cite{Rinchiuso:2020skh} strongly motivates determining the existing IACT sensitivity.

For these reasons, in addition to a broad model-independent consideration of final states, we will evaluate the IACT sensitivity to the Higgsino, Wino, and Quintuplet.
As these are full models, we can completely specify the particle physics contribution to Eq.~\eqref{eq:dmflux}.
Indeed, if we fix $m_{\DM}$ to the thermal masses mentioned above, then these models have no free parameters at all (up to a choice of mass splittings for the Higgsino, discussed below), although it is interesting to consider a broader range of masses in case the early Universe departed from the minimal thermal relic cosmology.
For each of these WIMPs, annihilation to a two-photon final state is possible, making a gamma-ray line at the DM mass a key target.
Yet a full determination of the cross-section and gamma-ray yield in these scenarios requires accounting for a number of effects, including Sommerfeld enhancement, resummation of effects of order $m_{\DM}/\mW$, and additional channels beyond the direct annihilation to two-photons, including endpoint photons, continuum emission, and the production and decay of bound states.
For the Wino, all these effects have now been included (for details see Refs.~\cite{Baumgart:2014vma,Bauer:2014ula,Ovanesyan:2014fwa,Baumgart:2014saa,Ovanesyan:2016vkk,Asadi:2016ybp,Baumgart:2017nsr,Baumgart:2018yed,Beneke:2018ssm,Beneke:2019vhz}), and we use the next-to-leading logarithmic (NLL) computation of Ref.~\cite{Baumgart:2018yed}.
The same formalism has recently been extended to the Quintuplet, and we use the results from the soon to appear calculation of Ref.~\cite{quintuplet:upcoming}.
For both the Wino and Quintuplet, the spectrum includes line-like photons from the two-body decay, lower energy continuum photons (arising from final states such as $W^+W^-$, cf. Fig.~\ref{fig:GammaYield}), and also endpoint photons arising from final states $\gamma + X$, where $X$ has a small invariant mass, so that the photon is almost line-like.
An identical computation has not yet been performed for the Higgsino (although see Refs.~\cite{Baumgart:2015bpa,Beneke:2019gtg,Beneke:2022eci}), and so here we follow the approach in Ref.~\cite{Rinchiuso:2020skh} of including the leading order (LO) computation of the line and continuum, but with the inclusion of Sommerfeld enhancement.
Finally, the Higgsino has an additional model parameter we need to specify, which is the splitting between the charged and neutral states in the spectrum, denoted $\delta m_+$ and $\delta m_N$, respectively.
For this purpose, we will choose one of the benchmarks used in Refs.~\cite{Baumgart:2015bpa,Rinchiuso:2020skh}, in particular taking $\delta m_+ = 350$ MeV and $\delta m_N = 200$ keV, saturating the limits set by direct detection.
The impact of varying the splittings is not considerable, and was considered in Ref.~\cite{Rinchiuso:2020skh}.

\subsection{Dark-matter distribution in the Inner Galaxy}
\label{subsec:DM_distr}

\begin{figure*}[!ht]
\includegraphics[width=0.47\textwidth]{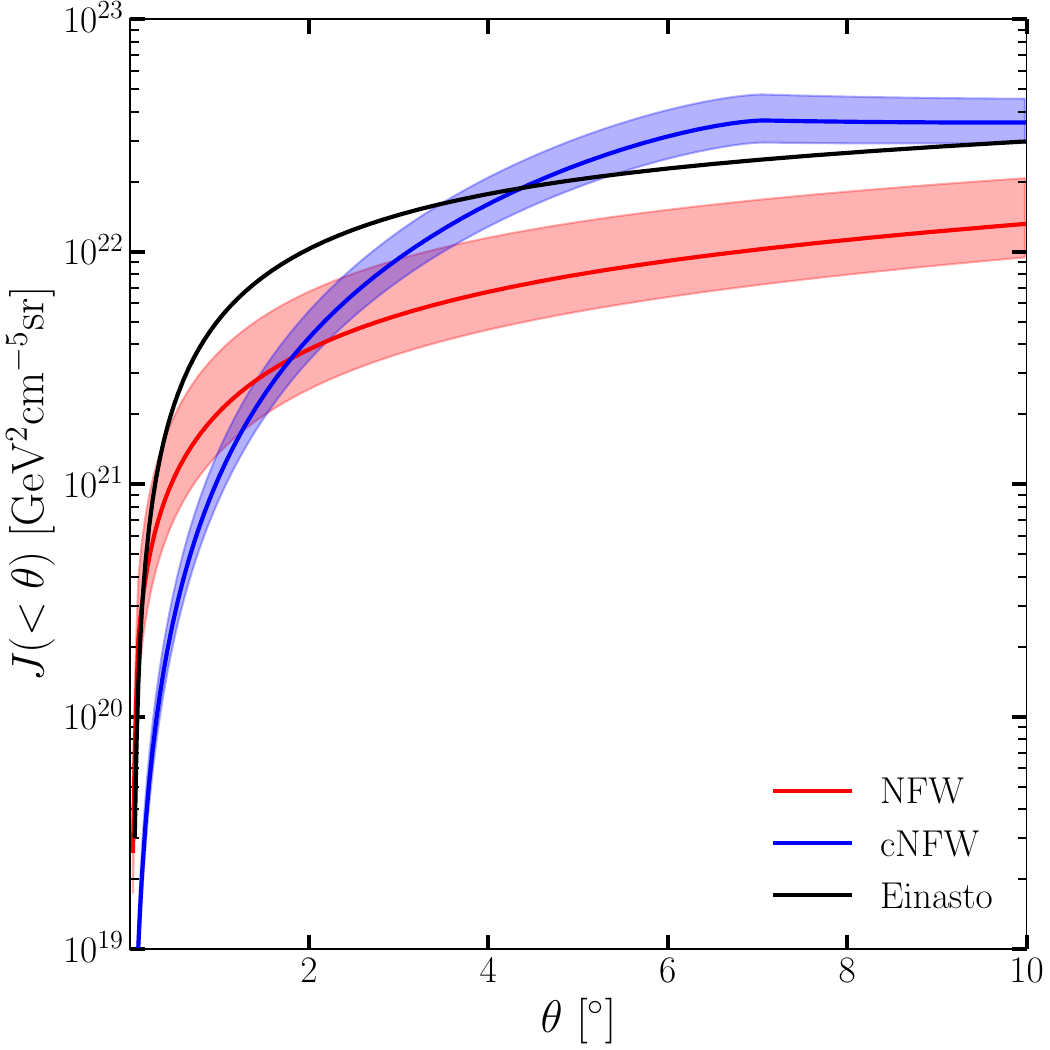}
\includegraphics[width=0.47\textwidth]{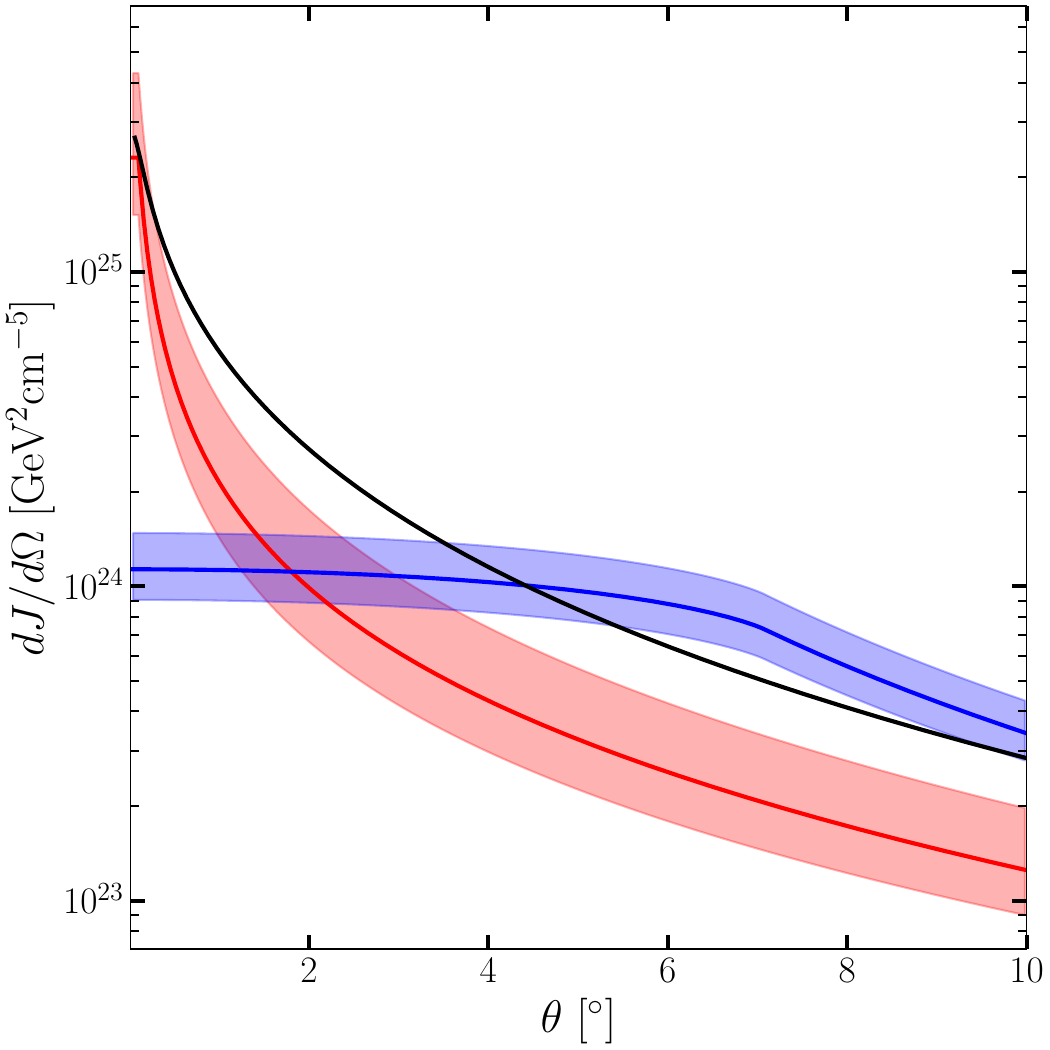}
\caption{Cumulative (left, as defined in Eq.~(\ref{eq:dmflux})) and differential (right) $J$-factor profiles, expressed in GeV$^2$cm$^{-5}$sr and GeV$^2$cm$^{-5}$, respectively, versus the angular distance from the GC, $\theta$.
Results are shown for three profiles, NFW (red line), cNFW (blue line), and Einasto (black). 
The red and blue-shaded regions correspond to the 1$\sigma$ uncertainty band for the NFW and cNFW profile parametrizations, respectively.
We will use the difference between these profiles to estimate the impact of systematic uncertainties in the Milky Way DM profile on the IACT indirect detection sensitivity.}
\label{fig:Jfacvstheta}
\vspace{-0.5cm}
\end{figure*}

The final ingredient required to determine the gamma-ray flux from DM annihilation is the DM distribution in the inner galaxy.
Inferring $\rho_\DM(r)$ in the central region of the Milky Way, however, is particularly challenging, and so in addition to direct measurements from modeling the mass using kinematic measurements of the gravitational potential, our understanding is conventionally complemented by cosmological simulations of structure formation.
DM-only simulations (see, for instance, Refs.~\cite{Springel:2008by,Diemand:2008in,10.1111/j.1745-3933.2009.00699.x}) predict cuspy DM distributions usually parameterized as NFW~\cite{Navarro:1996gj} or Einasto~\cite{Springel:2008by} profiles.
Once baryonic physics and feedback processes are included, the complexity increases dramatically (see, for example, Refs.~\cite{10.1093/mnras/stw145,10.1093/mnras/sty1690,2019MNRAS.490.4877P}), and this can lead to a modified DM profile.
For example kpc-sized cores can develop in the center of Milky Way-like galaxies~\cite{Mollitor:2014ara,Chan:2015tna} (although stellar measurements disfavor cores larger than several kpc~\cite{Hooper:2016ggc}).
Alternatively, feedback can lead to a contraction of the inner profile, significantly enhancing the DM density at the inner radii~\cite{Hopkins:2017ycn,McKeown:2021sob}.
To make direct measurements of $\rho_\DM(r)$ in our own Milky Way, one first has to infer the distribution of all matter via probes of the gravitational potential, and then subtract from this the baryonic contribution.
This latter step suffers from large uncertainties, which inevitably propagate to the derivation of the DM distribution (see, for example, Refs.~\cite{Iocco:2015xga,Portail:2016vei}).
For the above reasons, the DM distribution cannot be firmly established in the innermost region of the Milky Way, and this will generate a systematic uncertainty for DM searches via indirect detection.

In order to assess the impact of this systematic uncertainty, we will consider three different choices for $\rho_\DM(r)$.\footnote{An alternative approach is to take a given profile, and study the impact of sequentially larger core sizes, see Refs.~\cite{Rinchiuso:2018ajn,Rinchiuso:2020skh}.} 
For one of these profiles, we make use of a determination of the Milky Way mass profile using Gaia DR2 measurements of the rotation curve together with in-depth modeling of the baryonic components in the GC~\cite{Cautun:2019eaf}.
The authors of Ref.~\cite{Cautun:2019eaf} show that the profile inferred for the DM distribution shows evidence of being contracted by the presence of baryons.
In particular, the profile is contracted with respect to the standard NFW profile, which is parametrized as
\begin{equation}
\label{eq:NFWprofile}
\rho_\NFW(r) \propto \frac{1}{(r/r_s)(1+r/r_s)^2}.
\end{equation}
where $r_s$ is the scale radius, and the profile can then be normalized by ensuring it has the correct density in the Solar neighborhood, $\rho_\odot$.
The resulting model they provide is non-parametric, although given its contracted nature we refer to it as a contracted NFW (cNFW) going forward (although we emphasize that the cNFW does not follow the form given in Eq.~\eqref{eq:NFWprofile}).
This approach does not allow for a measurement of the distribution within the inner 1 kpc of the inner Galaxy, and so to be conservative we assume that the density is cored within this radius.
Specifically, we take $\rho_\cNFW(r) = \rho_\cNFW(r_c)$ for $r \le r_c = 1~{\rm kpc}$.
We emphasize that given the present uncertainty on $\rho_\DM(r)$, we largely focus on conservative forecasts.
The DM density may well continue to increase towards the GC, rather than taking on a core, and indeed this would also be consistent with several of the halos observed in the FIRE-2 simulation~\cite{Hopkins:2017ycn,McKeown:2021sob}.
If such an increase of $\rho_\DM(r)$ at small radii turns out to be correct, then this can significantly enhance the expected DM annihilation flux, and the expected sensitivity of H.E.S.S. and future IACTs.
(See, for example, Ref.~\cite{Dessert:2022evk} for a demonstration of the enhancement these profiles can have on a Fermi-LAT data analysis.)

In addition to the cNFW profile, we also consider a conventional NFW profile using the best fit parameters determined in Ref.~\cite{Cautun:2019eaf}.
Lastly, for comparison with the previous results of Refs.~\cite{Abdallah:2016ygi,Abdallah:2018qtu}, we also consider an Einasto profile for the DM distribution with the parametrization taken from Ref.~\cite{Springel:2008by} given by
\begin{equation}
\label{eq:Einastoprofile}
\rho_\text{Einasto}  \propto \exp\left[-\frac{2}{\alpha} \left( \left(\frac{r}{r_s}\right)^\alpha - 1 \right)\right]\!,
\end{equation}
with $\alpha = 0.17$ and $r_s=20$ kpc.
The Einasto profile is normalized to the local DM density $\rho_\odot$ such that 
$\rho_\Einasto(r_\odot) = \rho_\odot$ = 0.39 GeV/cm$^3$~\cite{Catena:2009mf}.

Although we will not use it in our default analysis, when comparing the IACT sensitivity for two-body neutrino annihilation channels to those determined from the ANTARES telescope in Ref.~\cite{Albert:2016emp}, to make the comparison more direct we will adopt the NFW parameters used in that work, and label them as aNFW.
The various halo parameters we use are collected in Tab.~\ref{tab:profileparameters}.

Once the form of $\rho_\DM(r)$ has been specified, we can then compute the associated $J$-factors following Eq.~\eqref{eq:dmflux}.
In Fig.~\ref{fig:Jfacvstheta} we show the total integrated and differential per solid angle $J$-factors as a function of angle from the GC, denoted $J(<\theta)$ and $dJ/d\Omega$, respectively.
For the NFW and cNFW profiles, we use the results of Ref.~\cite{Cautun:2019eaf} to show the associated 1$\sigma$ uncertainties on the $J$-factors.
As can be seen, there is considerable variation in both the distribution and total density between the profiles.
At the most central radii the Einasto profile generally has the largest densities, but by the edge of the region we will analyze at $\theta = 4^{\circ}$, the integrated $J$-factor in the Einasto and cNFW profile are comparable.

\renewcommand{\arraystretch}{1.1}
\begin{table}
    \centering
    \begin{tabular}{ w{c}{2.5cm} | w{c}{1.5cm} | w{c}{1.5cm} | w{c}{1.5cm}}
    \hline
    \hline
    Profiles & NFW & cNFW & aNFW \\
    \hline
    $\rho_\odot$ [GeV/cm$^3$]& 0.32 & 0.34 & 0.47\\
    $r_s$ [kpc] & 15.5 &  23.8 & 16.1\\
   \hline
   \hline
    \end{tabular}
    \caption{The local density $\rho_\odot$ and scale radius $r_s$ for the NFW, cNFW and aNFW profiles adopted in this work.
    The NFW and cNFW profiles are derived from Ref.~\cite{Cautun:2019eaf}, and we emphasize that the cNFW is non-parametric: it does not follow the form in Eq.~\eqref{eq:NFWprofile}.
    The aNFW profile is taken from Ref.~\cite{Albert:2016emp}, and used to compare with the results of that work.
    }
    \label{tab:profileparameters}
    \vspace{-0.5cm}
\end{table}
\renewcommand{\arraystretch}{1.1}
\begin{table*}[ht!]
\centering
\begin{tabular}{ w{c}{1.5cm} | w{c}{2.3cm} | w{c}{2.3cm} | w{c}{2.7cm} | w{c}{1.3cm} | w{c}{1.3cm} | w{c}{1.3cm} }
\hline
\hline
$i^{\rm th}$ ROI & Inner radius & Outer radius & Solid angle $\Delta\Omega$&\multicolumn{3}{c}{$J$-factor $J(\Delta\Omega)$}  \\
 &  [deg.] &  [deg.] & [10$^{-4}$ sr] &\multicolumn{3}{c}{[10$^{20}$ GeV$^2$cm$^{-5}$sr]}  \\
\hline
& & & & Einasto & NFW & cNFW\\
\hline
1 & 0.3 & 0.4 & 0.30 & 3.86 & 1.38 & 0.25 \\ 
2 & 0.4 & 0.5 & 0.50 & 5.28 & 1.86 & 0.43 \\
3 & 0.5 & 0.6 & 0.69 & 6.30 & 2.24 & 0.63 \\
4 & 0.6 & 0.7 & 0.88 & 7.06 & 2.51 & 0.84 \\
5 & 0.7 & 0.8 & 1.07 & 7.63 & 2.69 & 1.05 \\
6 & 0.8 & 0.9 & 1.26 & 8.07 & 2.83 & 1.26 \\
7 & 0.9 & 1.0 & 1.45 & 8.41 & 2.93 & 1.47 \\
... & ... & ... & ... & ... & ... & ... \\
37 & 3.9 & 4.0 & 7.55 & 8.82 & 3.32 & 7.81 \\ 
\hline
\hline
\end{tabular}
\caption{$J$-factor values in units of GeV$^2$cm$^{-5}$sr in each of the 37 annuli considered in this work for the three different DM profiles we consider.
We also show the exact boundaries of our regions, and their contained solid angles.
We emphasize that the solid angle and $J$-factor values account for the various masks we have imposed on our ROIs.
\label{tab:jfactors}}
\end{table*}

\subsection{TeV backgrounds in the Galactic Center}

Any signal of DM annihilation in the GC must be teased out from other background sources in that region.
In this subsection we discuss the dominant background contributions, and how we model them in our analysis.

The dominant background contribution arises from the combined flux of cosmic-ray hadrons, electrons, and positrons incident on the atmosphere, which is significantly larger than the observed rate for photons from even the brightest steady VHE sources.
Even though the showers that originate from charged cosmic-rays can be distinguished from gamma-ray showers, this discrimination is not perfect, and a residual contribution to the inferred photon flux is expected.
To account for this, we follow Ref.~\cite{2013APh43171B}, and compute the expected number of events produced by a flux of cosmic-ray hadrons (dominated by protons and helium nuclei), as well as electrons and positrons.
In particular, we assume a constant rejection power for protons and helium nuclei, whereas we assume no rejection for electrons and positrons.
The latter assumption is conservative, as the reconstruction of the primary interaction depth of the particle incident on the atmosphere can be used to help distinguish between gammas and electrons (see, for instance, Ref.~\cite{2009APh32231D}).
The misidentified cosmic-rays are then defined as the residual background.%
While for distances within $\lesssim 1$ kpc of the solar neighborhood, CR electron and positron sources may leave a spatial feature in the arrival directions of VHE cosmic rays, no anisotropy has been detected so far~\cite{Fermi-LAT:2017vjf}.
Therefore, the residual background is taken as spatially isotropic.

Beyond misidentified cosmic-rays, there will further be genuine photons incident on the atmosphere that have a non-DM origin and must be accounted for.
Indeed, the GC is a complex region populated by faint and diffuse photon emissions at TeV.
Among the conventional astrophysical sources are the H.E.S.S. Pevatron in the GC~\cite{Abramowski:2016mir}, Galactic diffuse emission~\cite{Ackermann:2012}, emission from the base of the Fermi Bubbles~\cite{Su_etal:2010,Fermi-LAT:2017opo}, and a possible contribution from a putative millisecond pulsar population in the Galactic bulge, see for instance, Ref.~\cite{Leane:2022bfm}.
Although we will account for it, we note that the emission of the GC Pevatron from current H.E.S.S. measurements is restricted within the inner $\sim$$75~{\rm pc}$ of the GC, which corresponds to an angular scale of $\sim$$0.5^\circ$~\cite{Abramowski:2016mir}.

The interactions of energetic cosmic-rays with interstellar material and ambient photon fields generates a diffuse flux of gamma rays known as the Galactic diffuse emission (GDE).
In particular, the GDE is generated by a combination of the photons arising from the decay of neutral pions produced from cosmic-ray proton collisions with interstellar gas, Bremsstrahlung from these same protons, and finally the inverse Compton scattering (ICS) of cosmic-ray electrons.
In the energy range accessible to Femi-LAT ($\sim {\rm MeV}-{\rm TeV}$) the GDE contributes the majority of photons detected by the telescope~\cite{Ackermann:2012}.
While the present uncertainty inherent in available GDE models represents a fundamental systematic error on many DM analyses performed with Fermi-LAT, this is not yet the case for H.E.S.S.~\cite{HESS:2022ygk}, where the GDE emission has yet to be conclusively detected.
However, as we look to deeper IACT sensitivities, the GDE can emerge as an important background contribution, and therefore we include a model for it in the present work.
Building realistic GDE models for CTA will be mandatory in order to maximize the utility of the increased flux sensitivity expected in the inner GC survey planned with the southern site of the CTA observatory; this point is discussed in Refs.~\cite{Silverwood:2014yza,Lefranc:2015pza,Moulin:2017cgb,Rinchiuso:2020skh}. 
As our purpose is simply to model a possible contribution from the GDE in a mock analysis, rather than confront real data, a simplified model of the GDE will suffice.
In particular, we make use of the ``GDE scenario 2'' developed in Ref.~\cite{Rinchiuso:2020skh}.
The spatial distribution of the $\pi^0$ and Bremsstrahlung emission is assumed to trace the morphology of interstellar dust, as determined in Ref.~\cite{Schlegel:1997yv}, and a simple parametric model for the ICS is adopted from Ref.~\cite{Su_etal:2010}.
For both, the energy distribution is then fitted to the spectrum measured by Fermi-LAT in Ref.~\cite{Fermi-LAT:2017opo}.
For the complete details, we refer to Ref.~\cite{Rinchiuso:2020skh}.

The Fermi Bubbles (FB) are giant double-lobe structures extending significantly above and below the Galactic plane -- out to $\sim$$55^\circ$ -- that were discovered using the Fermi-LAT satellite~\cite{Su:2010qj}.
At Galactic latitudes higher than $10^\circ$, the FB emission exhibits a power-law energy-spectrum, scaling as $E^{-2}$, although the spectrum softens considerably above 100 GeV.
At latitudes closer to the Galactic plane, Refs.~\cite{Acero:2016qlg,Fermi-LAT:2017opo,Storm:2017arh,Herold:2019pei} have detected brighter and harder emission using Fermi-LAT data, in particular with a power-law spectrum that persists until $\sim$$1~{\rm TeV}$.
However, the limited photon statistics above 100 GeV collected by Fermi-LAT obstruct any strong claims regarding a possible softening of the spectrum at higher energies.
Recent observations carried out by H.E.S.S. at the base of the FBs show no significant signal above 1 TeV, which requires a significant softening of FB emission spectrum in the TeV energy range~\cite{Moulin:2021mug}.
In order to model the FB emission in our ROI, we adopt the best-fit spectrum above 100 GeV from Ref.~\cite{Moulin:2021mug}.
For the spatial distribution, we assume an energy independent morphology, following the template derived in Ref.~\cite{Herold:2019pei}.

The final background contribution we consider is associated with the excess of gamma rays emerging from the GC that has been detected by Fermi-LAT~\cite{Goodenough:2009gk,Hooper:2010mq,Fermi-LAT:2015sau}, commonly referred to as the GCE.
While the nature of the excess remains debated (see Ref.~\cite{Leane:2022bfm} for a recent discussion), it may originate from a population of millisecond pulsars (MSP) in the inner galaxy.
Should this be the case, then electrons accelerated in the wind regions or magnetospheres of pulsars could escape the pulsar environment and undergo ICS on ambient radiation fields to produce VHE gamma rays, thereby generating an additional ICS background to our analyses.
The spectral index of the injection spectrum of $e^{\pm}$ from pulsars is difficult to constrain.
The most energetic $e^{\pm}$ arise from magnetic reconnection in the equatorial current sheet outside the pulsar light cylinder~\cite{Cerutti:2015hvk}, while the pulsed emission is expected to be generated in the polar cap region close to the pulsar magnetosphere.
The maximum energy of the emitted $e^{\pm}$ is also uncertain and could reach PeV energies~\cite{Guepin:2019fjb}.
With these caveats in mind, we have adopted the emission spectrum presented in Ref.~\cite{Macias:2021boz}, which can be roughly represented as a power-law spectrum $E^{-2.5}$ that is cut-off exponentially at 1 TeV.
For the spatial morphology, we have adopted the Boxy Bulge distribution as described in Ref.~\cite{Macias:2021boz}.
Although there are considerable uncertainties on both the spectrum and morphology of the MSP contribution, in addition to the ongoing debate as to its existence, we will find that the contribution has only a minor impact on IACT DM analyses, and therefore these uncertainties will not propagate through to the searches of interest.

\section{Dark-matter Search Strategy and mock data analysis}
\label{sec:analysis}

In the previous section we detailed the physical mechanisms underlying the emission components we will consider.
Here we describe the various experimental and observational considerations required to convert these into mock datasets, and define our strategy for detecting within that data a putative DM signal on top of the background photon emission.

\subsection{Forecasting the ultimate H.E.S.S.-like \\ dark-matter observation}
\label{subsec:forecastHESS}
In order to assess the ultimate sensitivity of current IACTs to DM in the GC region, we construct mock datasets that realistically build upon the most powerful existing analyses.
For the case of H.E.S.S. the deepest sensitivities were achieved in Ref.~\cite{HESS:2022ygk}, where observations of the innermost $\sim$$3^{\circ}$ of the Milky Way were constructed over six years, yielding a total observation time of 546 hours.
Prior to this, the most sensitive DM analyses, such as Ref.~\cite{Abdallah:2016ygi}, were constructed from observations taken along the Galactic plane, whose primary focus had been the detection of astrophysical sources in the plane and in the Central Molecular Zone, rather than maximizing the sensitivity to a DM signal.
Here we consider a search strategy that looks to build upon existing analyses by collecting observations out to larger latitudes, where the DM signal is expected to continue growing (see Fig.~\ref{fig:Jfacvstheta}) and where for certain DM profiles, greater separation from the background emission components can be expected.\footnote{The optimal search strategy will depend upon the assumed DM distribution.
For distributions highly peaked near the GC, additional observations near the dynamic center of the Milky Way will likely be optimal, as opposed to the strategy we pursue here.
Resolving the ideal strategy for different profiles will be an important target for upcoming H.E.S.S. observations, although we do not seek to fully resolve that here.}

An important question is the feasibility of extending existing analyses to include observations at higher latitudes.
With a location near the tropic of Capricorn in the southern hemisphere, the H.E.S.S. observatory is well suited to observe the central region of the Milky Way under favorable observational conditions.  
In particular, the GC can be observed from March to September each year with zenith angles lower than $30^\circ$.
Observations with lower zenith angle are generally preferred as they allow for lower energy thresholds and improved control of observational systematic uncertainties, given the thinner layer of atmosphere along the line of sight.
We note, however, that larger zenith-angle observations do allow for enhanced sensitivity to higher incident energy photons.
For reference, observations performed at a zenith angle of $\sim$$50^{\circ}$ have an energy threshold of $\sim$$1~{\rm TeV}$ and potentially higher systematic uncertainties.
Further discussion of these points can be found in Ref.~\cite{Adams:2021kwm}.

The right-ascension band of GC visibility contains a broad range of astrophysical objects.
Accounting for the prioritization of different objects in such a band, at most 150 hours of observations near the GC can be realistically obtained per year~\cite{Visibilty}. 
We assume a mean zenith angle of $20^{\circ}$ for the observations as an appropriate value considering the various constraints in this visibility window.
In detail, we assume data collected with stereo observations using the full five-telescopes array CT1-5, and use the appropriate instrument response functions extracted from Ref.~\cite{holler2015photon}.
These choices impose an accessible energy range between 200 GeV and 70 TeV~\cite{HESS:2022ygk}.

\begin{figure*}[!t]
\includegraphics[width=0.47\textwidth]{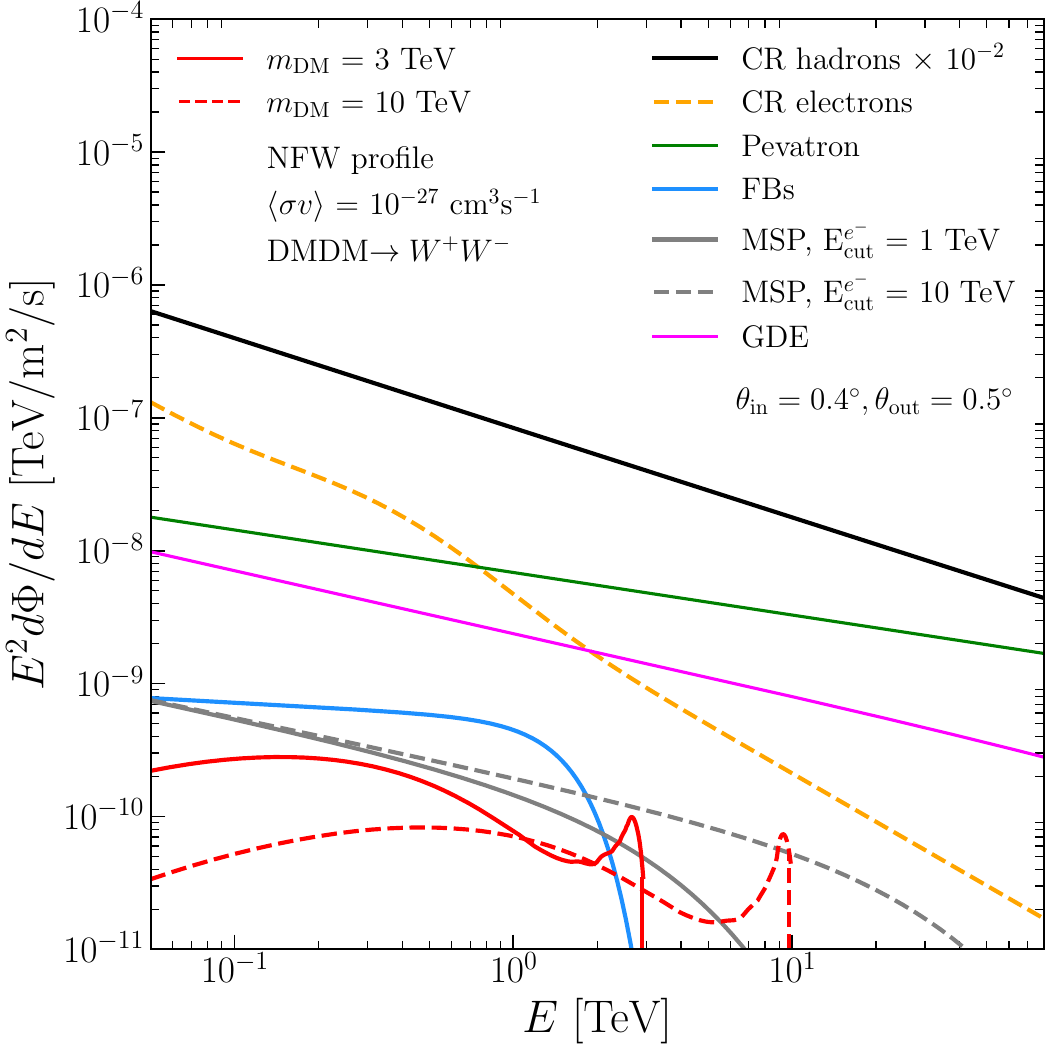}
\hspace{0.5cm}
\includegraphics[width=0.47\textwidth]{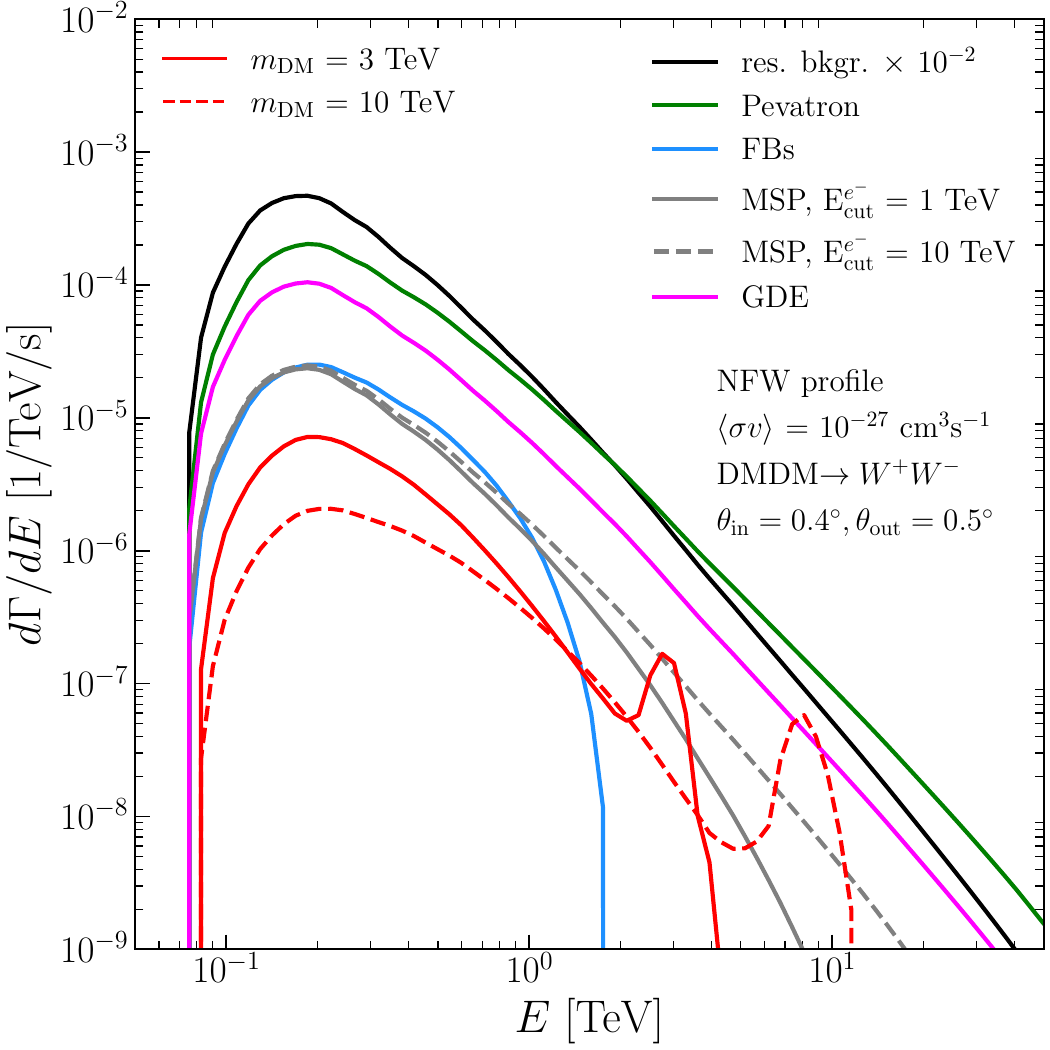}
\caption{{\it Left panel:} DM and background gamma-ray fluxes expected in the region from $0.4-0.5^{\circ}$ of the GC (our ROI 2).
For DM, we show the spectra for the self-annihilation of WIMPs of masses $m_{\DM}$ = 3 TeV and 10 TeV, respectively, in the $W^+W^-$ annihilation channel and with a velocity-weighted annihilation cross section $\langle \sigma v \rangle = 10^{-27}$ cm$^3$s$^{-1}$.
The hadronic (proton + helium)  (solid black line) and electron (orange line) cosmic-ray fluxes are shown.
The former is the largest background, and indeed in the plot we show the spectra at 1\% of the expected value, and it remains the dominant contribution.
The diffuse fluxes from the H.E.S.S. Pevatron~\cite{Abramowski:2016mir} (green line), the base of the Fermi Bubbles~\cite{Moulin:2021mug} (blue line) together with expectation from the MSP-bulge population~\cite{Macias:2021boz} (gray lines) assuming a power-law electron spectrum with an energy cut-off of 1 and 10 TeV, respectively, are shown.
Finally, we show a model for the Galactic Diffuse Emission (pink line), which is the sum of the $\pi^0$, bremsstrahlung, and ICS components.
{\it Right panel:} Energy-differential count rates as a function of energy for the same signal and backgrounds in the same region.
As opposed to the left hand plot, here the results have been convolved with the relevant instrumental effects.
Residual background is shown as the sum of the rates for CR hadrons and electrons. 
Note that the Pevatron emission is restricted to the inner 0.5$^\circ$ of the GC~\cite{Abramowski:2016mir}.}
\label{fig:SpectraDMbkgr}
\vspace{-0.5cm}
\end{figure*}

Combining these various considerations, we consider a flat exposure time of 500 hours distributed evenly across the inner $4^{\circ}$ of the GC to be an achievable target for H.E.S.S. before the advent of CTA.
As such, these are the parameters we adopt in our analysis, which will be sufficiently representative for us to interrogate the possible limiting factors, such as systematic uncertainties, on the ultimate IACT reach.
Nevertheless, we point out that this is likely a conservative assumption as to the ultimate DM dataset H.E.S.S. can collect.
Recently, in Ref.~\cite{HESS:2022ygk}, H.E.S.S. has performed a DM analysis in the inner $3^{\circ}$ of the GC using 546 hours of observations collected during phase 2 of the instrument, so an analysis with a larger amount of data over a smaller region than we consider (and we will directly contrast our results to that work).
Further, phase 1 of H.E.S.S. collected 254 hours of data in the inner $1^{\circ}$, which was used for DM searches in Refs.~\cite{Abdallah:2016ygi,Abdallah:2018qtu}.
Taken together, H.E.S.S. already has 800 hours of potential ON region data near the GC.
Combined with a collection rate for additional GC data of 150 hours per year, H.E.S.S. appears poised to amass over 1,000 hours of GC observations within the next two years.
However, the total exposure time is not distributed evenly across the inner $4^{\circ}$ of the GC. Therefore, as the  combination of inhomogeneous exposures and different instrument response functions is beyond the scope of this work,  we will focus only on 500 hours of ON region observations in the main body of this work. Still, we emphasize once more that this implies our results should be interpreted as conservative. We show, nevertheless, what impact of assuming 1,000 hours of flat exposure across the ON region could have on the sensitivity in Appendix~\ref{sec:appendixC}.
Finally, we will assume that the analysis can exploit the conventional ON and OFF method, where the ON region is as described above, and by design has an enhanced DM flux.\footnote{An alternative to the ON-OFF approach is to use Monte Carlo simulations to predict the residual background for the given observational and instrumental conditions~\cite{Holler:2020duc}.}
The OFF region is used as a control dataset to constrain the dominant background: charged cosmic rays.
The dataset is collected under the same instrumental and observational conditions as for the ON region, thereby allowing a realistic determination of the residual background in the signal region.
Traditionally, the OFF data is generated from observations taken close to the ON region.
However, if the OFF region is too close to the GC then it will also receive a contribution from DM annihilation and the other emission sources in inner Galaxy.
Given this, in the present work we will assume the OFF region is generated from extragalactic observations, sufficiently far from the GC that it effectively only contains photons arising from misidentified cosmic rays.
H.E.S.S. has collected many extragalactic observations, and amongst these datasets will exist that have been collected under comparable conditions to the ON region; indeed, exactly this approach has already been exploited in Ref.~\cite{Abdallah:2016ygi}.

\subsection{Defining and dividing the region of interest}
\label{subsec:ROISearch}

We next outline how to divide up our initial region of interest (ROI): an observation collected over the inner $4^{\circ}$ of the GC.\footnote{We note that taking an ROI that is a disk centered on the dynamical center of the Milky Way has been a common approach in H.E.S.S. DM searches, see Refs.~\cite{Abdallah:2016ygi,Abdallah:2018qtu,Rinchiuso:2018ajn}.}
In order to exploit the spatial characteristics of the DM signal with respect to the background, our ON region ROI is further divided into 37 annuli of width $\Delta \theta = 0.1^{\circ}$, and defined concentrically with inner radii ranging from $\theta_i = 0.3^{\circ}$ up to $\theta_i = 3.9^{\circ}$.

As was already mentioned, the GC is a crowded region containing many astrophysical VHE sources.
In order to avoid the challenges and systematic uncertainties associated with the modeling of the complex astrophysical mechanisms associated with this region, we deliberately exclude a fraction of the full disk.
In particular, we mask a box with Galactic longitudes $|l|<1^{\circ}$ and latitudes $|b|<0.3^{\circ}$ in order to mask sources in the Galactic plane, which explains why our innermost annuli only begins at $0.3^{\circ}$.
We further exclude a disk of radius $0.8^\circ$ centered at $(l, b) = (-1.29^{\circ}, -0.64^{\circ})$ to cover the bright source HESS J1745-303.
This procedure certainly does not exclude all background contributions from our ROI, and the remaining backgrounds are instead modeled, as described in Sec.~\ref{subsec:signalbackground}.

In Tab.~\ref{tab:jfactors}, we provide further details of our ROI, including the solid angle and the corresponding $J$-factors obtained in each annulus, excluding the masked regions that overlap with the ROI.
In practice, the $J$-factors are computed for all the pixels of the ROI, and then the total $J$-factor of each ROI is obtained by summing all the values in the pixels that are not lying in the masked regions.
Further, the left panel of Fig.~\ref{fig:SpectraDMbkgr} shows the expected fluxes in our second ROI, including the expected signal generates by DM of mass $m_\DM = 3~{\rm TeV}$ or $10~{\rm TeV}$ and annihilating into $W^+ W^-$ with $\langle \sigma v \rangle = 10^{-27}~{\rm cm}^3{\rm s}^{-1}$, for two different DM profiles (NFW and cNFW).
In addition to DM, the relevant instrumental and astrophysical contributions are shown.

\subsection{Expected number of DM and background \\ photons in the GC region}
\label{subsec:signalbackground}

For a given DM annihilation channel and density profile, we need to compute the expected number of signal events, $N^S_{ij}$, in the $i^{\rm th}$ ROI and $j^{\rm th}$ energy bin.
If we denote the $i^{\rm th}$ ROI as having solid angle $\Delta\Omega_i$, and the $j^{\rm th}$ energy bin as having width $\Delta E_j$, then the expected counts is given by
\begin{widetext}
\begin{equation}
    N^S_{ij} = T_{{\rm obs}, i} \int_{E_j - \Delta E_j /2}^{E_j + \Delta E_j /2} dE \int_{-\infty}^{\infty} dE'\, \frac{d\Phi^S_{ij}}{dE'} (\Delta \Omega_i, E')\, A_{\rm eff}^{\gamma}(E')\, G(E - E').
\label{eq:count}
\end{equation}
\end{widetext}
Here $E'$ and $E$ are the true and reconstructed energies, respectively, $d\Phi^S_{ij}/dE'$ is the predicted DM flux as given in Eq.~\eqref{eq:dmflux}, $T_{{\rm obs}, i}$ is the time $i^{\rm th}$ ROI was observed, $A_{\rm eff}^\gamma$ is the energy-dependent photon effective-area, and $G$ corresponds to the finite energy resolution of the instrument.
We model the energy resolution as a Gaussian with width specified by $\sigma/E = 10\%$, following Ref.~\cite{2009APh32231D}.
The energy resolution is a critical parameter when considering searches for DM that involve a two-photon final state, as is the case for the Wino, Higgsino, and Quintuplet.
Improvements in the energy resolution allow the associated line-like features to be more easily extracted from the smoother background components, and for instance at TeV energies, CTA is expected to achieve energy resolutions as low as 5\%~\cite{CTAperformances}.
The gamma-ray acceptance used for five-telescope H.E.S.S. observations of the inner halo of the Milky Way is extracted from Ref.~\cite{holler2015photon}, which enables us to obtain a realistic description of the instrument response for  the survey we consider.
Further refinements in the description of the instrument response would require dedicated simulations of the instrument and observations, and could further improve our estimated sensitivity, but is beyond the scope of the present work.
We adopt a homogeneous exposure over all the ROIs resulting in a total of  500 hours of high-quality observations.
Given the limited field of view of IACTs, such an exposure requires a dedicated observation strategy with the definition of a grid of pointing positions.
For a possible implementation of such a strategy, see Ref.~\cite{HESS:2022ygk}, however we will not pursue the design of an optimized strategy here.

A similar procedure can be used to generate the expected number of background and cosmic-ray events, $N^B_{ij}$ and $N^{\rm CR}_{ij}$, respectively.
For the background, we use Eq.~\eqref{eq:count}, but with the following substitution
\begin{equation}
    \frac{d\Phi^S_{ij}}{dE}\, A_{\rm eff}^{\gamma}
    \to
    \frac{d\Phi^B_{ij}}{dE}\, A_{\rm eff}^{\gamma}
\end{equation}
where $d\Phi^B_{ij}/dE$ is the background photon flux.
A similar procedure holds for the cosmic rays, up to the different acceptances for the hadronic and leptonic cases.
In detail, we now substitute
\begin{equation}
    \frac{d\Phi^S_{ij}}{dE}\, A_{\rm eff}^{\gamma}
    \to
    \frac{d\Phi^{p,{\rm He}}_{ij}}{dE}\, A_{\rm eff}^{{\rm CR}}
    + \frac{d\Phi^{e^+e^-}_{ij}}{dE}\, A_{\rm eff}^{\gamma}.
\end{equation}
The hadronic cosmic-ray contribution experiences a different effective area, which can be expressed in terms of the photon acceptance as $A^{\rm CR}_{\rm eff} = \epsilon^{\rm CR} A^{\gamma}_{\rm eff}$ where $\epsilon^{\rm CR}$ is the cosmic-ray rejection efficiency.
While the majority of hadronic cosmic rays can be efficiently rejected, a fraction of them will be misidentified as photons, and here we assume a representative value of $\epsilon^{\rm CR}=10\%$ over the full energy range.
This allows us to reach a photon efficiency of higher than 95\%~\cite{2009APh32231D}.
As mentioned earlier, we assume that the showers originating from electrons and positrons cannot be discriminated from photons, and therefore include their flux in $d\Phi^B_{ij}/dE$.

Following Eq.~\eqref{eq:count}, the differential count rate for the emission in each bin is straightforwardly obtained by
\begin{equation}
    \frac{d\Gamma^X_{ij}}{dE} = \frac{1}{T_{{\rm obs}, i}} \frac{dN^X_{ij}}{dE},
    \label{eq:rates}
\end{equation}
with $X=S$, $B$, or ${\rm CR}$.
The right panel of Fig.~\ref{fig:SpectraDMbkgr} shows the expected gamma-ray rates in our second ROI from a DM annihilation signal, misidentified cosmic-rays, and the above-mentioned conventional astrophysical emissions: the Pevatron, FBs, MSPs, and the GDE for ROI 2.

\subsection{Analysis method}
\label{subsec:Statmethod}

A likelihood-ratio test statistic technique is used to compute the IACT sensitivity.
We use a two-dimensional Poisson likelihood, indexed by the spatial ROI and the energy bin $i$ and $j$.
In particular, we combine the statistically independent measurements in the ON and OFF region into the following combined likelihood,
\begin{equation}\begin{aligned}
    &\mathcal{L}_{ij} (N_{ij}^S, N_{ij}^B,N_{ij}^{\rm CR}, \bar{N}_{ij}^S, \bar{N}_{ij}^B, \beta_{ij}\, |\, N_{ij}^{\rm ON}, N_{ij}^{\rm OFF}) \\ 
    =\,\, &\textrm{Pois}[\beta_{ij} (N^S_{,ij} + N^B_{ij}+N^{\rm CR}_{ij}), N^{\rm ON}_{ij}] \\
    \times\,\, &\textrm{Pois}[\beta_{ij}(\bar{N}^S_{ij} + \bar{N}^B_{ij} + \alpha_i N^{\rm CR}_{ij}), N^{\rm OFF}_{ij}] \\
    \times\,\, &\frac{1}{\sqrt{2\pi} \sigma_{\beta}} \exp \left[ - \frac{(1-\beta_{ij})^2}{2\sigma_{\beta}^2} \right]\!.
    \label{eq:likelihood_beta}
\end{aligned}\end{equation}
Here $\textrm{Pois}[\lambda,n] = e^{-\lambda}\lambda^n/n!$, and the number of observed counts in each region and bin is given by $N^{\rm ON}_{ij}$ and $N^{\rm OFF}_{ij}$.
The expected number of counts in the signal and background regions is given by $\beta_{ij} (N^S_{ij} + N^B_{ij}+N^{\rm CR}_{ij})$ and $\beta_{ij}(\bar{N}^S_{ij} + \bar{N}^B_{ij} + \alpha_i N^{\rm CR}_{ij})$; assuming the datasets were collected under sufficiently similar conditions, the cosmic-ray contribution will only differ by the different region sizes, $\alpha_i \equiv \Delta\Omega_i^{\rm OFF}/\Delta\Omega_i^{\rm ON}$.
The expected counts in the ON region included the contributions from a possible DM signal, the residual background cosmic-rays, the diffuse emission, Fermi bubbles, MSPs, and also in the inner two ROIs, the Pevatron.
Under our assumptions that the OFF region is an identical observation far from the GC, we take $\bar{N}_{ij}^S = \bar{N}_{ij}^B = 0$, and $\alpha_i=1$.

Finally, we include two nuisance parameters to test the impact of various systematic uncertainties, although we will not include these effects in our default analyses.
The first is encapsulated in the factor $\beta_{ij}$, which is a Gaussian nuisance parameter used to account for observational and instrumental systematic uncertainties~\cite{Silverwood:2014yza, Lefranc:2015pza, Moulin:2019oyc}.
It is applied as a normalization factor to the expected number of events, and in order to account for a systematic uncertainty of 1\% on the measured event rates we take a flat value of $\sigma_{\beta} = 0.01$ (following Ref.~\cite{HESS:2022ygk}).
In principle, one could allow the nuisance parameter to vary across the different bins in a controlled manner, although we have not pursued that possibility here.
In practice, in each bin we can remove the nuisance parameter with the profile-likelihood technique.
The second nuisance parameter we will append to Eq.~\eqref{eq:likelihood_beta} is associated with the uncertainty on the $J$-factor which is central to determining the DM flux.
We model this systematic uncertainty through the use of a log-normal distribution of mean $\bar{J}$ and width $\sigma_J$ (see, for instance, Refs.~\cite{Lisanti:2017qoz,HESS:2020zwn,HESS:2021zzm}), expressed as
\begin{equation}\begin{aligned}
    \mathcal{L}^J (J | \Bar{J}, \sigma_J) =\, &\frac{1}{\ln(10) J \sqrt{2\pi} \sigma_J}  \\
    \times\, &\exp \left[-\frac{( \log_{10}J - \log_{10}\bar{J})^2}{2\sigma^2_J} \right]\!.
    \label{eq:likelihood_Jfac}
\end{aligned}\end{equation}
We multiply the likelihood in Eq.~\eqref{eq:likelihood_beta} by this factor, and given all the other parameters determine $J$ again with the profile likelihood technique.
We will consider the impact of this systematic for the NFW and cNFW profiles, using the central values and uncertainties shown in Fig.~\ref{fig:Jfacvstheta}.
(We will not consider this effect for the Einasto profile.)
Again this systematic effect will be tested, but not included in our results by default.

\begin{figure*}[!ht]
\includegraphics[width=0.47\textwidth]{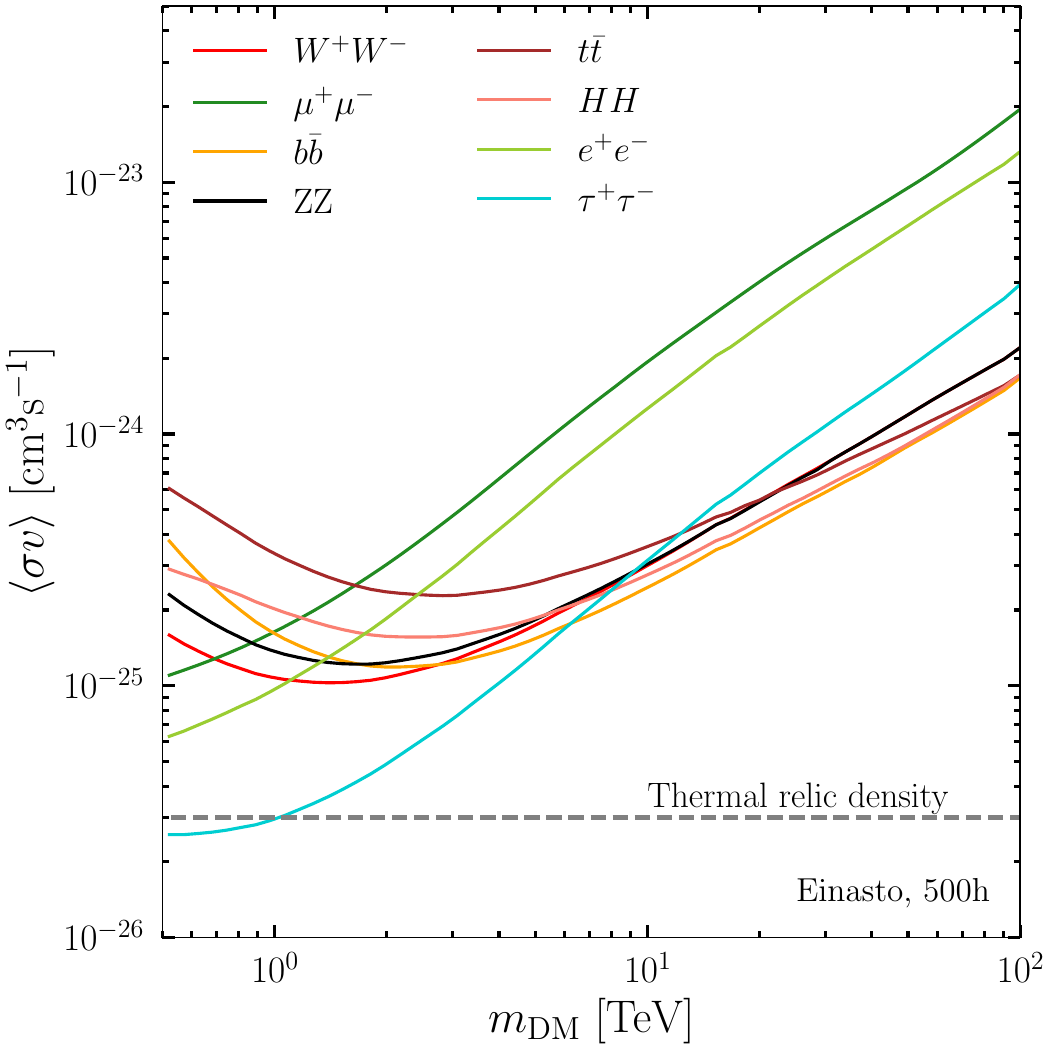}
\includegraphics[width=0.47\textwidth]{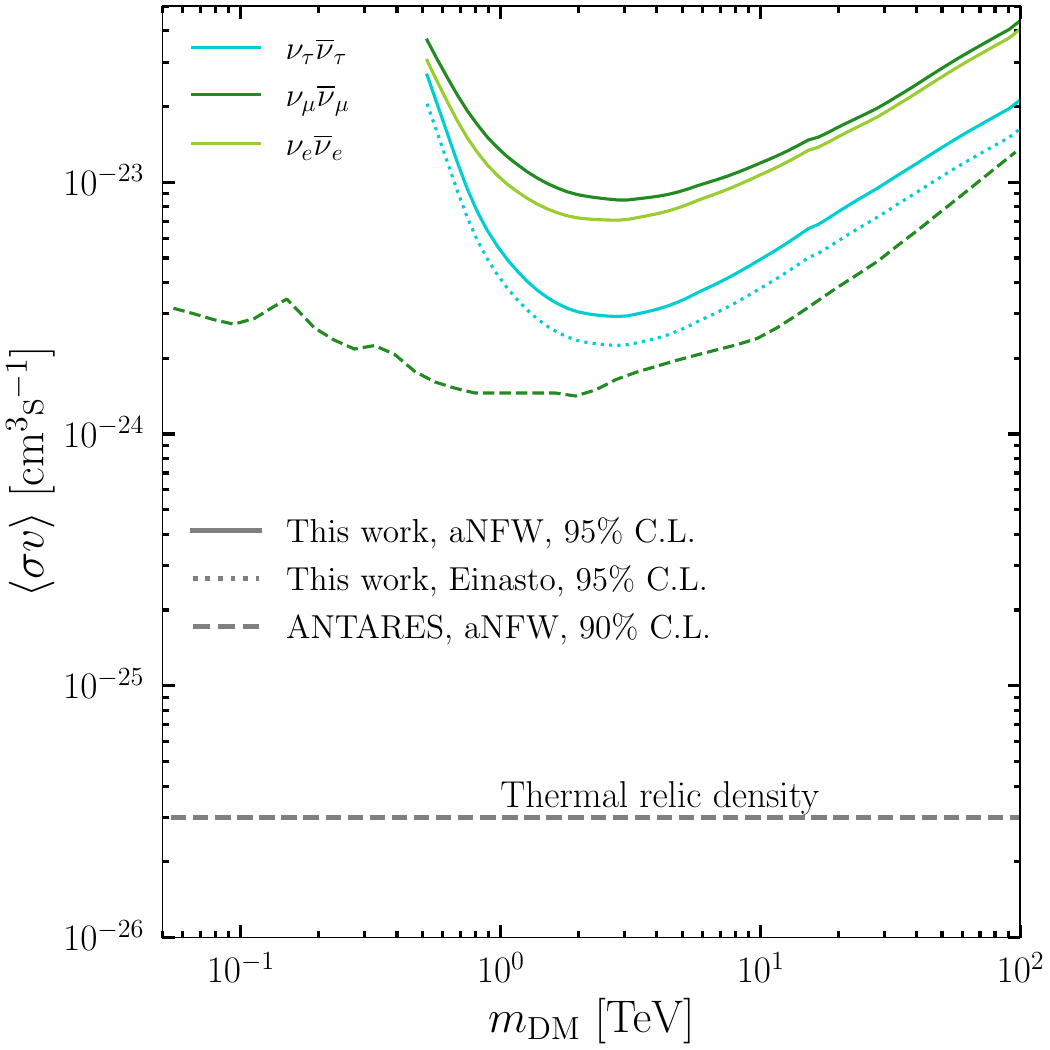}
\caption{{\it Left panel:} 
Mean expected 95\% upper C.L. limits on $\langle \sigma v \rangle$ as a function of the DM mass $m_{\DM}$ for six different two-body final states.
All spectra are computed with \texttt{HDMSpectra}, and we show results for the assumption of the DM distributed in the inner galaxy as the Einasto profile.
The dashed gray horizontal line represents the expected cross section for a conventional thermal relic.
{\it Right panel:} Equivalent results for neutrino final states, and the equivalent sensitivity obtained by ANTARES for the $\nu_{\mu} \bar{\nu}_{\mu}$ channel~\cite{Albert:2016emp} (although the limits in that work are 90\% C.L.).
We emphasize that although the prompt annihilation is to $\nu \bar{\nu}$, H.E.S.S. can constrain this channel as the electroweak corrections will generate photons at the considered masses.
To facilitate the comparison, we adopted the NFW parameters used in Ref.~\cite{Albert:2016emp}, which we label as the aNFW profile.
The limits that we obtain for the $\nu_{\tau} \bar{\nu}_{\tau}$ channel are also shown for the assumption of DM distributed according to the Einasto profile, as adopted for the left panel.}
\label{fig:LimitsChannels}
\vspace{-0.5cm}
\end{figure*}

The full likelihood is then the product of the $J$-factor weighted likelihood in Eq.~\eqref{eq:likelihood_beta} over all spatial and energy bins.
Within the combined likelihood, all remaining nuisance parameters, such as the rate of cosmic rays or the background normalizations, are removed with the profile likelihood.
Once this is done, the likelihood depends solely on the particle DM properties, which for a given spectrum, are specified by $\langle \sigma v \rangle$ and $m_\DM$.
Imagining we fix the mass, the likelihood can then be used to define a test statistic (TS) for setting upper limits~\cite{Cowan:2011an},
\begin{equation}
    \text{TS}(m_{\DM}) = - 2 \ln \frac{\mathcal{L}(\langle \sigma v \rangle, m_{\DM})}{\mathcal{L}(\widehat{\langle \sigma v \rangle}, m_{\DM})},
    \label{eq:TS}
\end{equation}
where $\widehat{\langle \sigma v \rangle}$ denotes the value of the cross section which maximizes the likelihood for a given $m_{\DM}$.
In the limit of large statistics, this TS follows a $\chi^2$ distribution with a single degree of freedom.
Assuming we are working in this limit, we can set one-sided 95\% upper limits on $\langle \sigma v \rangle$ by solving for the cross section above the best fit where ${\rm TS} = 2.71$, and this is the procedure we use to compute our limits given a mock dataset.

The expected limits, and confidence intervals on these, can be determined by applying the above procedure to a large set of Monte Carlo datasets, and then analyzing the distribution of limits obtained from these.
(This approach has been widely used by H.E.S.S., see, for instance, Refs.~\cite{Abdallah:2016ygi,Abdallah:2018qtu}.)
An alternative approach is to follow the Asimov procedure of Ref.~\cite{Cowan:2011an}.
Instead of generating many realizations of the expected background, we simply take the mean expected background as the data to compute the mean of the expected sensitivity.
The Asimov approach can also be used to compute confidence intervals of the expected sensitivity~\cite{Cowan:2011an}.
In particular, the $N$-sigma containment band can be computed from ${\rm TS} = (\Phi^{-1}[0.95] \pm N)^2$, where $\Phi^{-1}$ is the inverse of the cumulative distribution function of a normal distribution with $\mu=0$ and $\sigma=1$.
We will use the Asimov procedure throughout this work, although we validate in App.~\ref{sec:appendixA} that good agreement is achieved with a conventional Monte Carlo approach (we find that the mean expected limit and 1$\sigma$ containment bands agree within 5\% and 4\%, respectively).
Finally, we power constrain our limits so that we do not allow them to move below the expected one-sigma lower limit~\cite{Cowan:2011an}.

\section{Results}
\label{sec:results}

Here we combine the formalism introduced so far to estimate the ultimate H.E.S.S.-like sensitivity to DM.
We begin by showing our results for two-body final states, and we will use these as a testing ground for the various systematic uncertainties we explore.
After this we will consider the reach to the three specific DM scenarios we consider: the Wino, Higgsino, and Quintuplet.

\subsection{Sensitivity to two-body final states}
\label{subsec:SensLim}

Firstly we consider limits on the model-independent approach where we search for DM annihilating into various two-body final states with the spectra determined from \texttt{HDMSpectra}.
For the various channels considered, we determine the H.E.S.S.-like sensitivity through the use of mean expected upper limits at 95\% C.L. on the annihilation cross section $\langle \sigma v \rangle$ as a function of the DM particle mass from 0.5 up to 100 TeV.
Note that the range of photon energies we consider is from 0.2 to 70 TeV, and is set by experimental and observational parameters we have adopted.

Our results are shown in Fig.~\ref{fig:LimitsChannels}.
In the left panel we depict the limits for the non-neutrino channels for the Einasto profile described in Sec.~\ref{subsec:DM_distr}.
For a mass of 1.5 TeV, the sensitivity reaches $1.0\times10^{-25}~{\rm cm}^3{\rm s}^{-1}$ and $3.4\times10^{-26}~{\rm cm}^3{\rm s}^{-1}$ for the $W^+W^-$ and $\tau^+\tau^-$ annihilation channels, respectively. 
At the lower masses we consider, H.E.S.S. is most sensitive to leptonic final states, which produce a large fraction of their photons through final-state radiation.
This occurs, as at the lowest masses the primary factor determining the limit is not the total fraction of $2m_{\DM}$ converted to photons, but rather how much of the photon energy is above the lower energy threshold, and final-state radiation produces a spectrum peaked near $m_{\DM}$.
At higher masses, as more of the generated photons enter the observable energies, the total energy converted into photons dictates the sensitivity, and hadronic channels become the most constrained.

In the right panel of Fig.~\ref{fig:LimitsChannels}, we show the sensitivity for two-body neutrino final states when adopting the NFW parameterizations of the Milky Way DM distribution used in Ref.~\cite{Albert:2016emp}, referred as to aNFW profile in Tab.~\ref{tab:profileparameters}.
For the $\nu_{\tau} \bar{\nu}_{\tau}$ channel, we further show results obtained for the assumption of DM distributed according to the Einasto profile, so that the difference between that and the aNFW can be seen.
These results are compared with the 90\% C.L. mean expected upper limits from ANTARES~\cite{Albert:2016emp} for the $\nu_{\mu} \bar{\nu}_{\mu}$ channel.
As anticipated for DM masses well above the weak scale, a H.E.S.S.-like IACT array is clearly sensitive to neutrino final states given the large number of VHE photons such final states can generate when electroweak corrections are incorporated.
Indeed, we see that IACTs are a competitive method to search for these channels.

\begin{figure}[!t]
\includegraphics[width=0.45\textwidth]{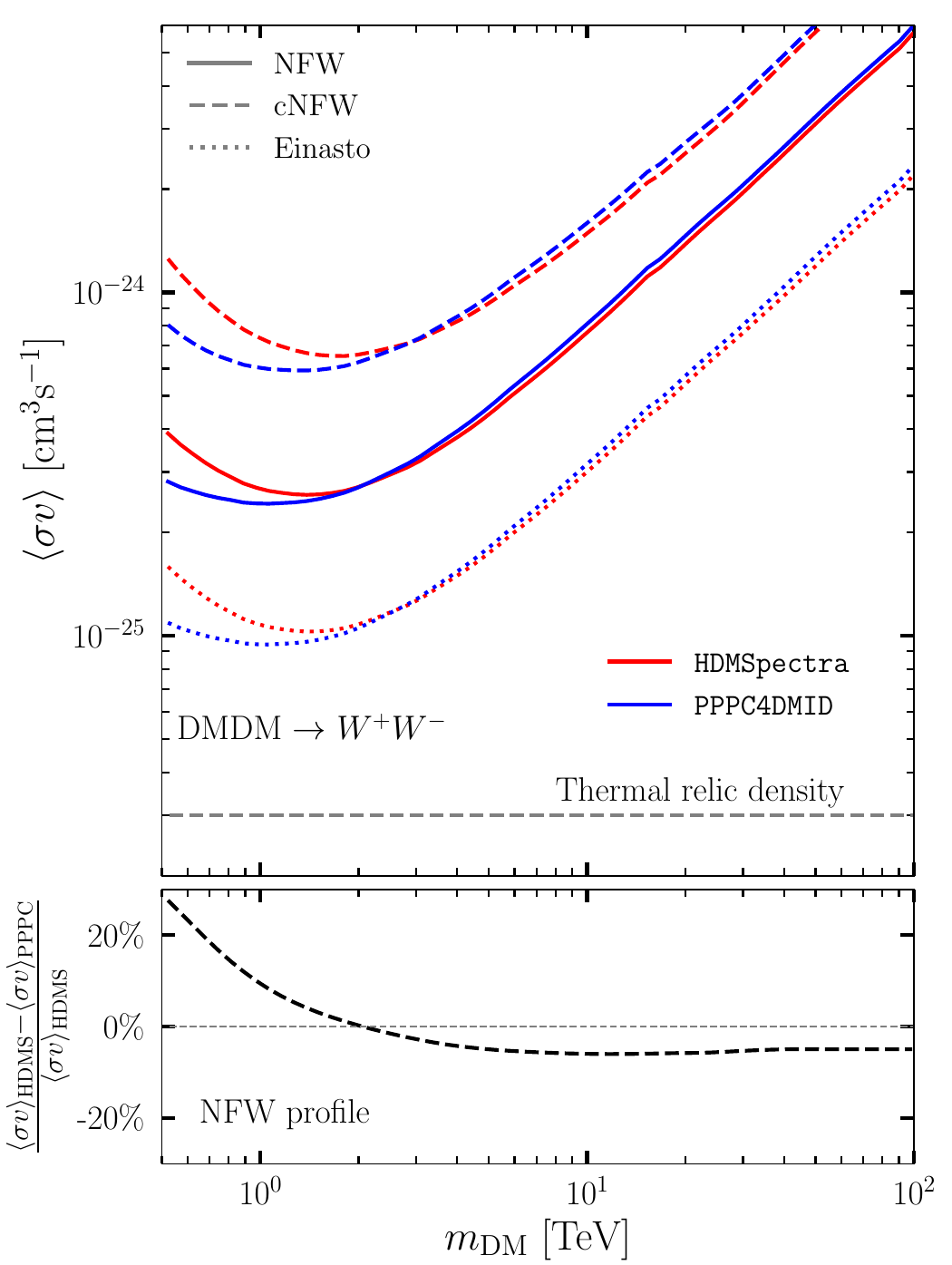}
\caption{The impact on our limits of the systematic uncertainty on the DM signal prediction, focusing on a representative final state, the $W^+ W^-$ channel (shown in Fig.~\ref{fig:LimitsChannels}).
In the top panel we show the limit for the three different DM profiles we consider, the NFW (solid lines), cNFW (dashed lines), and Einasto (dotted lines).
The variation between profiles can impact the limits by almost an order of magnitude.
In each case, we also show the limit obtained when using the spectrum as computed by \texttt{PPPC4DMID}~\cite{Cirelli:2010xx} (blue) and \texttt{HDMSpectra}~\cite{Bauer:2020jay} (red), cf. Fig.~\ref{fig:GammaYield}.
The impact of the spectrum is most pronounced at lower masses, and in the lower panel we show the percentage difference between the spectra for the NFW profile.
}
\label{fig:LimitsWWComparisonSpec}
\vspace{-0.5cm}
\end{figure}

We can also compare our results to those derived recently with actual H.E.S.S. observations in Ref.~\cite{HESS:2022ygk}.
A comparison of, for example, the $W^+ W^-$ channel at $m_\DM \sim1 {\rm TeV}$ shows that our expected sensitivity is weaker by a factor of $\sim$$2.5$.
This is not unexpected, and can be attributed to three key differences between the analyses.
The first and most significant difference arises from the updated energy-dependent photon effective-area adopted in Ref.~\cite{HESS:2022ygk}.
In this work, we use the publicly available instrument responses, however, improved ones compared to what we have assumed here were exploited in the recent analysis.
Secondly, we have assumed a 500 hour observation distributed uniformly over the inner $4^{\circ}$ (up to the regions we have masked).
In Ref.~\cite{HESS:2022ygk}, a 546 hour observation of the inner $3^{\circ}$ was used.
As the Einasto profile sharply peaks at inner radii (see Fig.~\ref{fig:Jfacvstheta}), observations at closer radii will enhance the expected sensitivity for this profile.
Moreover, the 546 hours used in Ref.~\cite{HESS:2022ygk} were not distributed uniformly over the region, but instead biased towards intermediate radii with an enhanced expected signal-to-noise ratio, which could lead to a further improvement in sensitivity.
Finally, that work used spectra from \texttt{PPPC4DMID}, which predicts a slightly higher photon yield than \texttt{HDMSpectra} (see Fig.~\ref{fig:GammaYield}, or Fig.~\ref{fig:LimitsWWComparisonSpec} and the discussion below).

\begin{figure}[!t]
\includegraphics[width=0.45\textwidth]{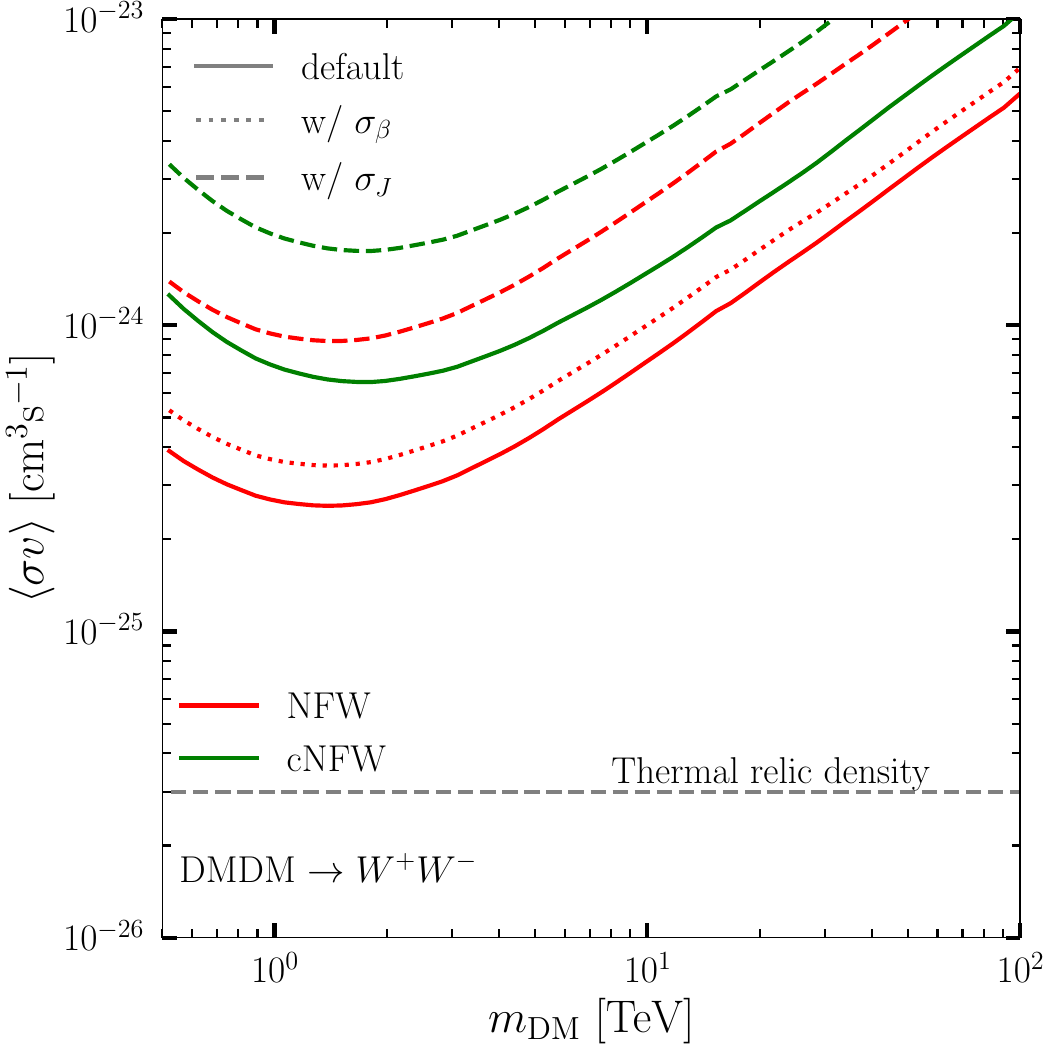}
\caption{The impact of our systematic nuisance parameters on the DM sensitivity.
For two different profiles, NFW and cNFW, we show the impact of profiling over the uncertainty on the $J$-factor as described in Eq.~\eqref{eq:likelihood_Jfac}, where $\sigma_J$ is determined from the width of the bands in Fig.~\ref{fig:Jfacvstheta}.
We further demonstrate the impact of our treatment of an additional systematic uncertainty encoded by $\beta_{ij}$ in Eq.~\eqref{eq:likelihood_beta}, labelled as $\sigma_{\beta}$, although we only show this for the NFW profile---the impact for the cNFW profile is comparable.}
\label{fig:LimitsJfactorBetaUncertainty}
\vspace{-0.5cm}
\end{figure}
\begin{figure}[!ht]
\centering
\includegraphics[width=0.45\textwidth]{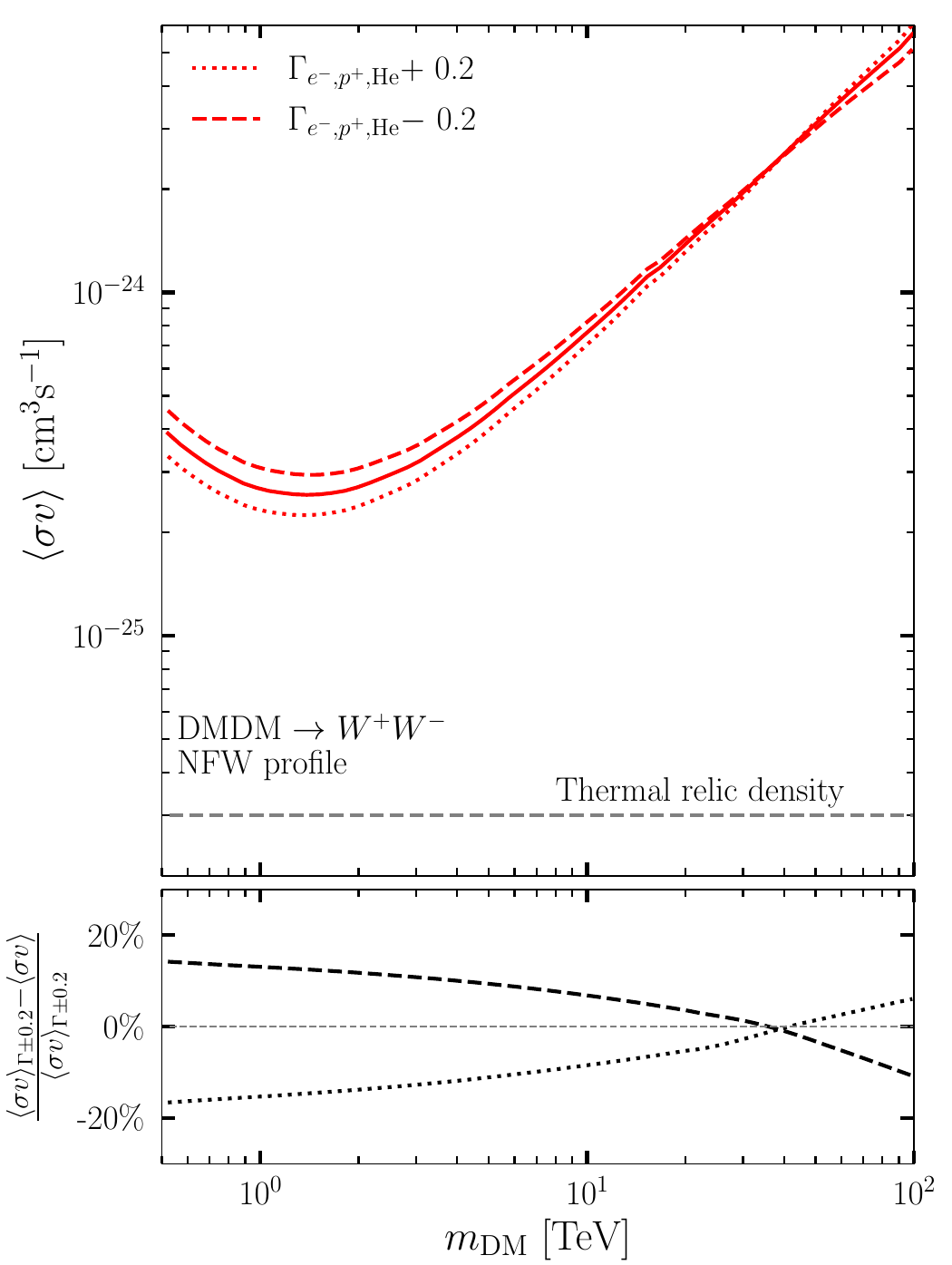}
\caption{{\it Top panel:} 95\% C. L.  sensitivity on $\langle \sigma v \rangle$ as a function of the DM mass $m_{\DM}$ for the  $W^+W^-$ channel and the NFW profile parametrization.
The horizontal grey long-dashed line is set to the value of the natural scale expected for the thermal-relic WIMPs. The dashed and dotted lines show the limits when the indeces of the power laws describing the spectra of cosmic rays are changed by $\pm$ 0.2. 
{\it Bottom panel:} percentage difference of the limits obtained for the two uncertainty values shown in the top panel and the limits with no uncertainty.}
\label{fig:LimitsWWComparisonBkgrMis}
\vspace{-0.5cm}
\end{figure}

\begin{figure}[!t]
\includegraphics[width=0.45\textwidth]{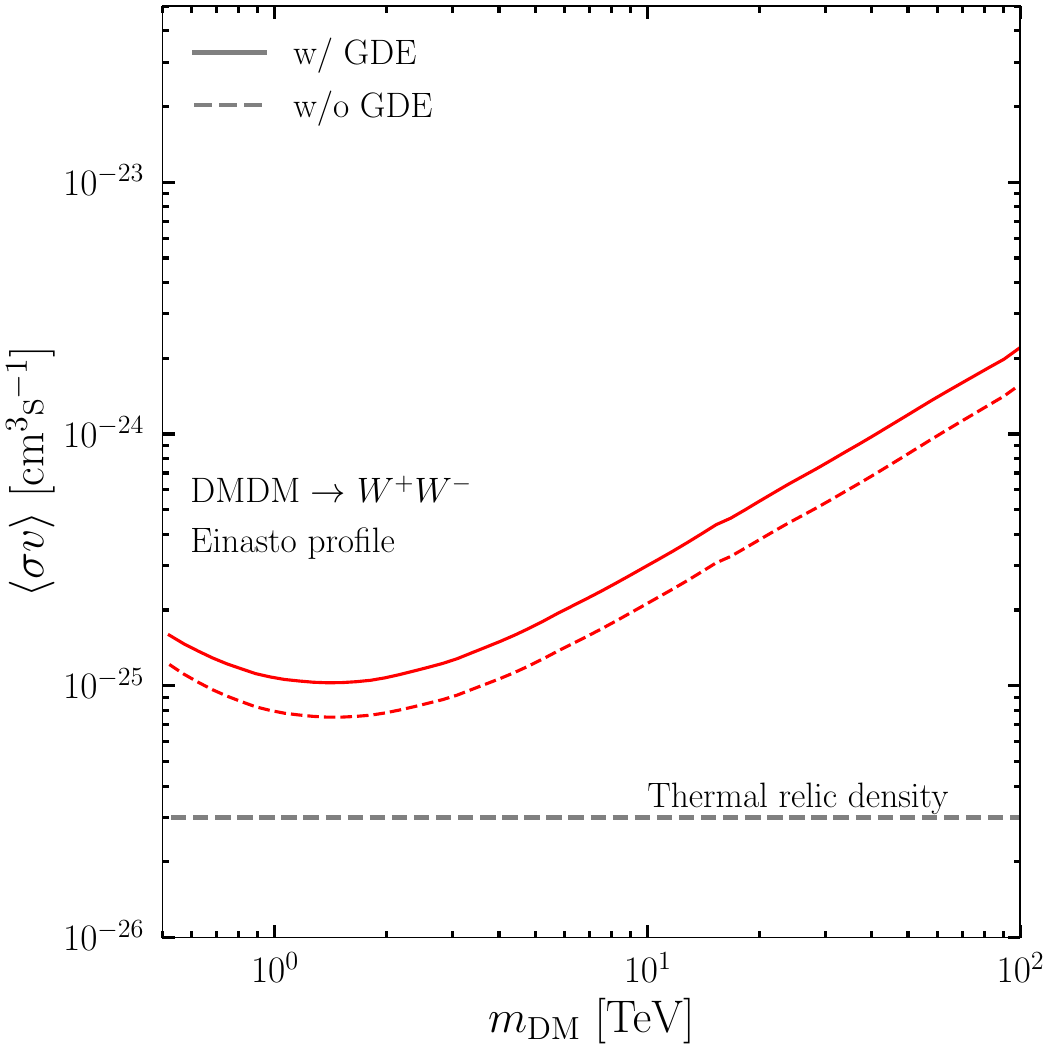}
\caption{Impact of the GDE contribution to the overall background on the 95\% C. L. sensitivity on $\langle \sigma v \rangle$ as a function of the DM mass. 
The DM distribution is assumed here to follow the Einasto profile and the DM particles self-annihilate into the   $W^+W^-$ channel.}
\label{fig:ComparisonGDEnoGDE}
\vspace{-0.5cm}
\end{figure}

\begin{figure*}[!ht]
\includegraphics[width=\textwidth]{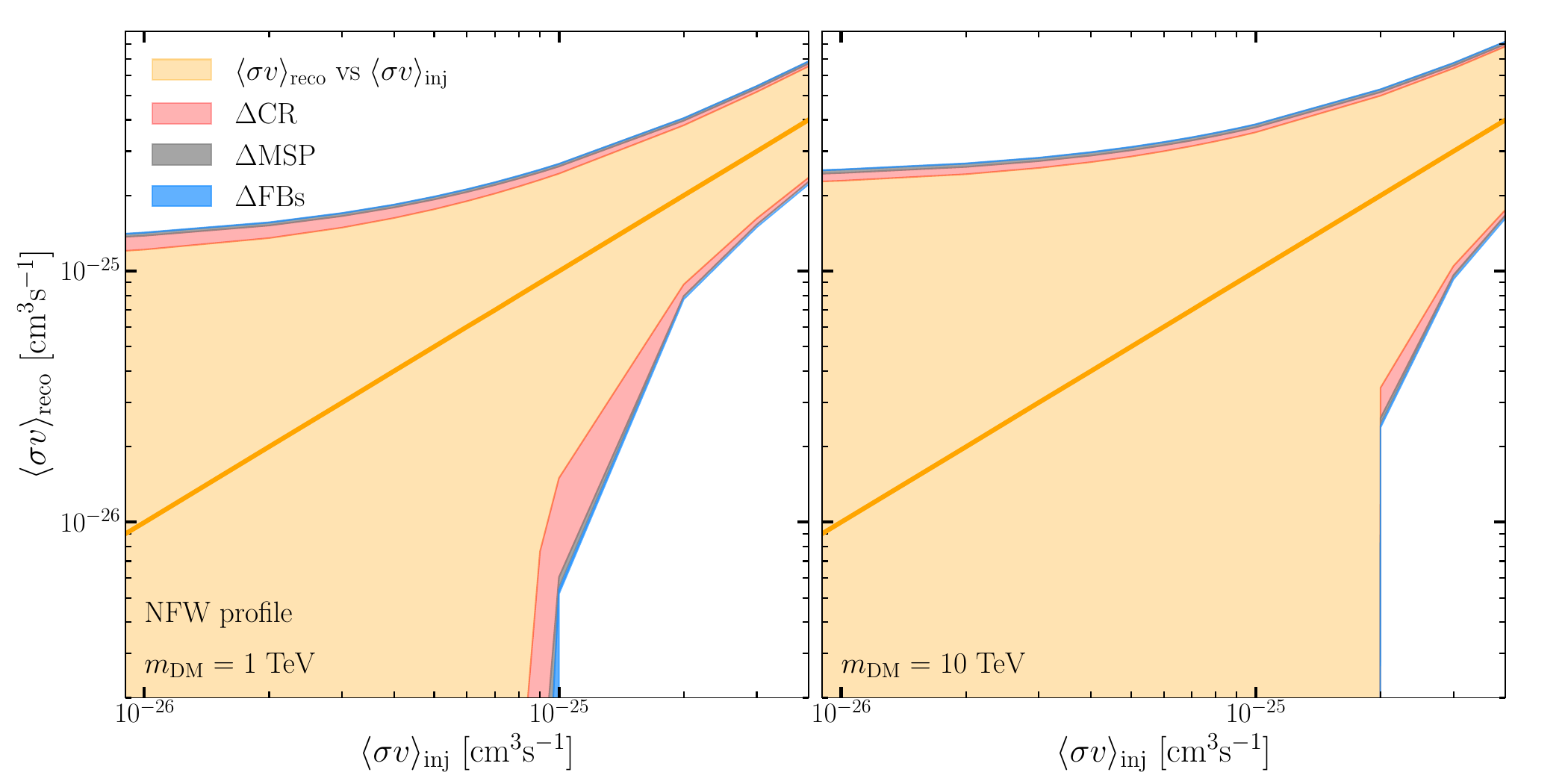}
\caption{The impact of background systematics on the ability to reconstruct an injected DM signal.
For two different DM masses -- 1 (left panel) and 10 TeV (right) -- and assuming annihilation to $W^+ W^-$ we inject a specific cross section, $\langle \sigma v \rangle_{\rm inj}$, and consider our ability to reconstruct this, labeled as $\langle \sigma v \rangle_{\rm reco}$.
The orange line and band show the mean and 1$\sigma$ reconstruction in the absence of background systematics.
If we add to this various background uncertainties, such as a variation in the cosmic-ray energy spectrum ($\Delta$CR), or similar variations to the MSP ($\Delta$MSP), and Fermi bubble ($\Delta$FB) contributions, they expand the uncertainty as shown in the red, gray, and blue bands, respectively.
(The uncertainties from each contribution are added in that order sequentially.)
See the text for further details.} 
\label{fig:RecoComparisonBkgrMis}
\end{figure*}

\subsection{Systematic uncertainties}
\label{subsec:SysLim}

We now take the two-body final states as a testing ground for the impact of systematic uncertainties on the ultimate H.E.S.S.-like sensitivity.
In particular, we will focus on annihilation to $W^+ W^-$ as a representative channel to study.
In this subsection, we will consider the impact of systematic uncertainties on the dark-matter signal, instrumental effects, and the backgrounds to dark matter.

\subsubsection{The dark-matter signal}

For a given mass and cross section, there are two aspects of the DM flux in Eq.~\eqref{eq:dmflux} subject to systematic uncertainty: our ability to compute the spectrum, $dN/dE$, for a given channel, and the impact of our incomplete knowledge of the DM distribution in the Milky Way on $J(\Delta \Omega)$ (see Sec.~\ref{subsec:SigSpec_thmodels}).
In Fig.~\ref{fig:LimitsWWComparisonSpec} we depict the impact of these uncertainties.

Firstly, we see the impact of varying between two possible computations of the spectrum, resulting from \texttt{PPPC4DMID}~\cite{Cirelli:2010xx} (blue lines) and \texttt{HDMSpectra}~\cite{Bauer:2020jay} (red lines).
As discussed in Sec.~\ref{subsec:SigSpec_thmodels}, differences between these two methods are expected to be pronounced when $m_\DM$ approaches the electroweak scale, or well above it.
The former concern clearly impacts the results: for the lowest masses we consider, the limits from \texttt{PPPC4DMID} are almost 30\% stronger than \texttt{HDMSpectra}.
Above $\sim$$1~{\rm TeV}$, however, the differences are much less pronounced, and converge to roughly 6\%.
This suggest that at 100 TeV, the highest mass at which \texttt{PPPC4DMID} provides results, the impact of effects resulting from multiple electroweak emissions are not significant, so the two approaches are in reasonable agreement.
(Note that \texttt{HDMSpectra} provides spectra for masses all the way to the Planck scale.)
This suggests that for DM searches in this mass range with IACTs, the focus should be on effects at the electroweak scale in order to reduce the theoretical uncertainty on the spectrum.
This, of course, should be pursued alongside improvements in the QCD uncertainties within \texttt{Pythia}, which were shown to vary between a few to fifty percent in Ref.~\cite{Amoroso:2018qga}.
We note that the $W^+ W^-$ channel is representative of this point across channels; to demonstrate this we show equivalent results for the $\mu^+ \mu^-$ channel in App.~\ref{sec:appendixB}.

While the uncertainties on the DM spectra are not negligible, the results in Fig.~\ref{fig:LimitsWWComparisonSpec} make manifest that the dominant uncertainty in the signal prediction arises from $J(\Delta \Omega)$, or specifically from the present uncertainties we have on the DM distribution in the Milky Way.
Given the narrow field-of-view of H.E.S.S., and IACTs more generally, even our extended ROIs are still primarily sensitive to the very inner part of the Milky Way DM profile.
With this in mind, even though the cNFW profile is suggestive of a DM distribution that is more sharply peaked towards the GC, given present observations cannot reliably probe the inner kpc, our conservative assumption to core the profile within this radius implies that the cNFW leads to the weakest sensitivity at present.
For instance, adopting the cNFW over the NFW profile degrades sensitivity by a factor from 1.8 up to 3.2, depending on the mass.
Further, as shown in Fig.~\ref{fig:LimitsJfactorBetaUncertainty}, when profiling over the uncertainty in the $J$-factor for a given profile -- using Eq.~\eqref{eq:likelihood_Jfac} -- the sensitivity degrades by a factor of 3.2 up to 3.6 and from 2.6 up to 2.7 across the probed mass range for the NFW and cNFW profiles, respectively.
This degradation is of a similar size to the impact of moving between profiles.
While challenging, any improvements in our understanding of $\rho_\DM(r)$ in the inner galaxy will immediately result in decreased uncertainties for DM analyses.

\subsubsection{Instrumental background}

As a proxy for various instrumental effects, and possibly even residual uncertainties in the background estimation, we incorporated a Gaussian nuisance parameter $\beta_{ij}$ into the likelihood in Eq.~\eqref{eq:likelihood_beta}.
As discussed already, we gave this parameter a flat (in space and energy) width of 1\%.
The impact of this parameter on our DM reach is shown in Fig.~\ref{fig:LimitsJfactorBetaUncertainty}, and can lead to a loss in cross-section sensitivity of between 1.2 to 1.4 across the masses we consider.

\subsubsection{Background mismodeling}

\renewcommand{\arraystretch}{1.1}
\begin{table*}[ht!]
\centering
\begin{tabular}{c|c|c|c|c}
\hline
\hline
\hspace{0.1cm}$m_{\DM}$\hspace{0.1cm} &$\langle\sigma v\rangle_{\rm inj}$&
\multicolumn{3}{c}{Uncertainty budget} \\
& & \hspace{0.1cm}Statistical\hspace{0.1cm}  & \hspace{0.1cm}Residual background\hspace{0.1cm} & \hspace{0.1cm}Conventional background\hspace{0.1cm} \\ 
TeV & \hspace{0.1cm}[cm$^3$s$^{-1}$]\hspace{0.1cm} & [\%] & [\%] & [\%]\\ 
\hline
\hline
1 & 2$\times$10$^{-25}$& 90\%  & 9\% & 1\% \\
\hline
10 &2$\times$10$^{-25}$& 94\%  & 5\% & 1\% \\
\hline
\hline
\end{tabular}
\caption{The uncertainty budget for the reconstructed annihilation cross section as determined from the analysis in Fig.~\ref{fig:RecoComparisonBkgrMis}.
The third column shows the statistical uncertainty obtained from the containment bands when no additional uncertainties on the background are included.
Fourth and fifth columns represent the uncertainties obtained when modifying the residual and conventional background emissions, respectively.
\label{tab:uncertainties}}
\end{table*}

Finally we explore the impact of systematics in the background modeling on the ultimate sensitivity to DM, considering possible sources from variations to the spectra to our background components.

As shown in Fig.~\ref{fig:SpectraDMbkgr}, even with a 90\% rejection efficiency, the dominant background remains hadronic cosmic-rays.
While the rejection efficiency can in principle be improved, a contribution from this background appears irreducible for any IACT DM search.
Given this, it is important that we know the spectral shape of this contribution as accurately as possible so that we can distinguish it from the predicted DM spectrum.
The uncertainty in the spectrum of cosmic rays reaching the Earth's upper atmosphere has been estimated by AMS-02~\cite{AMS:2015tnn}, where it was established that the spectral index of the proton flux is uncertain at the level of $\pm 0.2$, which we vary around our central value of $2.7$.
In Fig.~\ref{fig:LimitsWWComparisonBkgrMis} we demonstrate that this translates into an uncertainty on our limits of up to 17\%.
We similarly considered how our results varied with an energy cutoff to the PeVatron, a change in the spectrum index of the FBs or MSP spectra at the level of $\pm 0.2$, however in all these additional cases found no appreciable impact on our results.

\begin{figure*}[!t]
\centering
\includegraphics[width=0.88\textwidth]{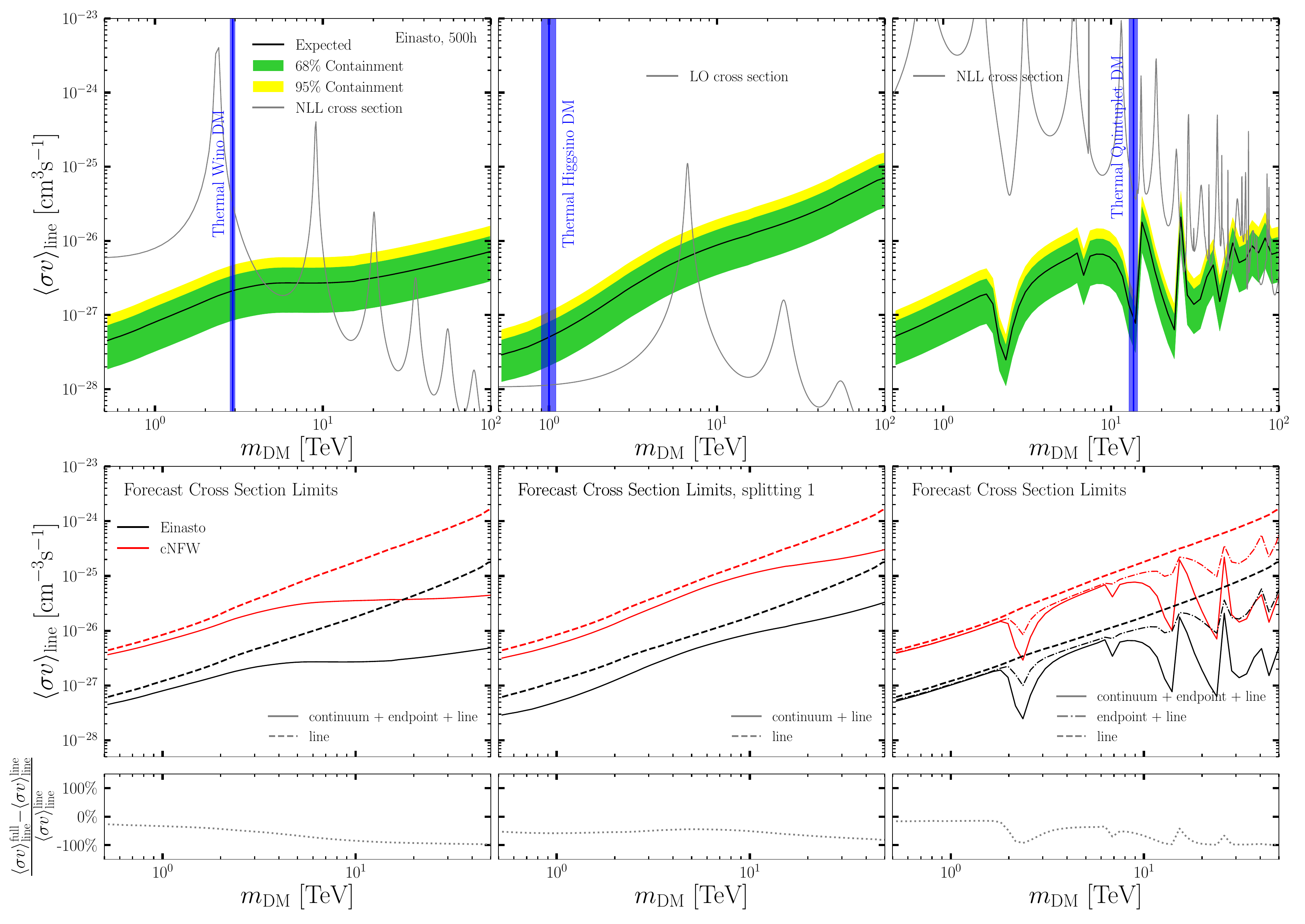}
\caption{Power-constrained mean expected 95\% limits on the line cross section for three canonical DM models: the Wino (left), Higgsino (middle), and Quintuplet (right).
In the top panels we show the sensitivity in each case (assuming an Einasto profile), which can then be compared to various theoretical predictions for the rates.
While these models can be considered for arbitrary masses, their is a unique $m_\DM$ signaled out as the mass where the correct relic abundance is obtained from a thermal cosmology, and these are labeled by the thermal vertical bands.
In the lower panels we show results for the cNFW halo, as well as a breakdown of the contribution to the limit from the line only in each case, and the other contributions such as the endpoint and continuum.
The sharp features in the quintuplet expected limit are physical, and explored in Fig.~\ref{fig:GammaYieldWinoHiggsinoQuintuplet} and in the text.
}
\label{fig:WinoHiggsinoQuintuplet}
\end{figure*}

To further explore the impact of background mismodeling, we perform a series of tests where we inject a known DM signal into the data and explore our ability to reconstruct it in the presence of systematic uncertainties.
The results of this analysis are shown in Fig.~\ref{fig:RecoComparisonBkgrMis}.
There we have fixed the DM mass (either 1 or 10 TeV), the annihilation channel (again $W^+ W^-$), and vary the injected cross section between $4 \times 10^{-25}~{\rm cm}^3{\rm s}^{-1}$ and $9 \times 10^{-27}~{\rm cm}^3{\rm s}^{-1}$.
For each value of the injected cross section, $\langle \sigma v \rangle_{\rm inj}$, we then compute the reconstructed annihilation cross section, $\langle \sigma v \rangle_{\rm reco}$, by maximizing the likelihood.
Firstly, in orange we show our ability to reconstruct the signal in the absence of background mismodeling.\footnote{Recall that we are using the Asimov procedure, so even for very small injected signals, our mean reconstructed cross section exactly matches the injected value.}
We see that for 10 TeV, our analysis can recover only values of $\langle \sigma v \rangle_{\rm inj} \geq 2\times10^{-25}~{\rm cm}^3{\rm s}^{-1}$; for lower $\langle \sigma v \rangle_{\rm inj}$ only upper limits can be obtained. 
These results immediately highlight that the statistical uncertainty on the recovery of the cross-section across the values we show is considerable.
We next consider how this is modified if we sequentially add various sources of background mismodeling.
Firstly, we vary the spectrum of the hadronic cosmic rays by $\pm 0.2$.
Next we vary the MSP and then FB spectra by the same value, each time adding the uncertainty from the additional contribution in quadrature.
While changes in the residual background has a visible effect on the ability to reconstruct an injected signal, the impact from the MSP and FB variations is negligible.
Nevertheless, in all cases the impact is considerably smaller than the initial statistical uncertainty.

\begin{figure*}[!t]
\centering
\includegraphics[width=0.88\textwidth]{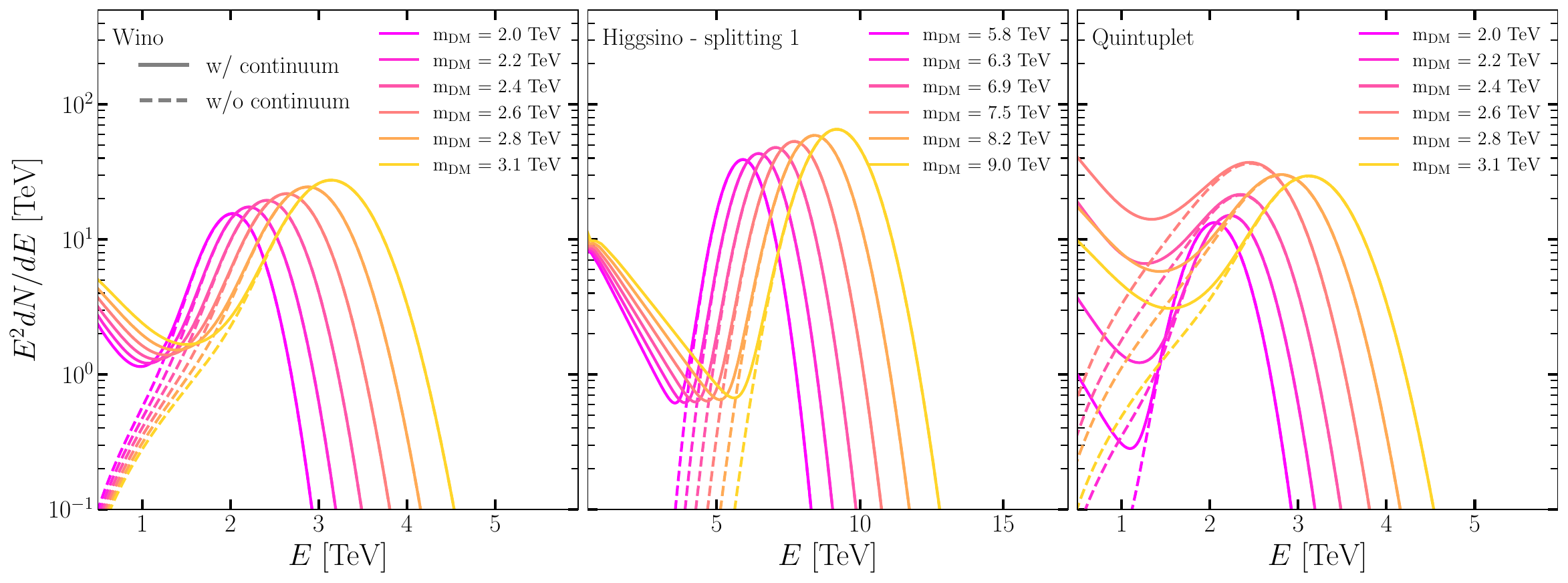}
\caption{The DM annihilation spectrum for the Wino (left), Higgsino (middle), and Quintuplet (right), with and without the continuum contribution added in, after convolution with the H.E.S.S. energy resolution.
Spectra with and without continuum are shown as solid and dashed lines, respectively.
(For the Wino and Quintuplet, the spectrum without the continuum includes the line and endpoint contributions, whereas the Higgsino includes only the line.)
The Wino and Higgsino spectra evolve smoothly as a function of mass, whereas the Quintuplet does not.
Locations where the Quintuplet spectrum evolves sharply give rise to the sharp variations in the limit as a function of mass seen in Fig.~\ref{fig:WinoHiggsinoQuintuplet}, and are discussed further in the text.}
\label{fig:GammaYieldWinoHiggsinoQuintuplet}
\vspace{-0.5cm}
\end{figure*}

Table~\ref{tab:uncertainties} summarizes the uncertainty budget for each case considered in this analysis.
The uncertainty on the reconstructed values $\langle \sigma v \rangle_{\rm reco}$ is largely dominated by the statistical uncertainty, indicating that the performance can still be improved with additional gamma-ray observations.
Indeed, the impact of limited statistics is the second largest uncertainty we have seen after the impact of the $J$-factor.
Relative to these, as already noted the residual background uncertainty has a slight impact on the achieved performance, whereas, the uncertainties on the MSP and FB astrophysical backgrounds can be considered as negligible.

In order to assess the impact of the GDE in the  background contribution, we show, in Fig.~\ref{fig:ComparisonGDEnoGDE}, 95\% C. L. sensitivity on $\langle \sigma v \rangle$ as a function of the DM mass $m_{\DM}$ computed with and without the inclusion of the GDE in the overall background budget following Eq.~(\ref{eq:likelihood_beta}).
The presence of the GDE contribution leads to a sensitivity loss between 1.3 and 1.4 across the masses considered.
This impact suggests that, with the current exposure, the GDE emission starts becoming an important background emission for H.E.S.S.-like DM searches in the GC region.
Such an emission could be within the reach of detection by H.E.S.S.-like IACTs at TeV energies.

\subsection{Sensitivity to Higgsino, Wino and Quintuplet dark-matter}
\label{subsec:canonicalmodels}

Next we apply our procedure to consider the canonical electroweak DM candidates of the Wino, Higgsino, and Quintuplet, as reviewed in Sec.~\ref{subsec:TeVmodels}.
The expected sensitivity to these scenarios is shown in the top panel of Fig.~\ref{fig:WinoHiggsinoQuintuplet}, where we have assumed the Einasto profile.
We show the 95\% C.L. mean expected upper limits together with the 1 and 2$\sigma$ containment bands.
In each case we also show the theoretical predictions for their cross sections.
Note in all cases, the cross section we consider is the weighting of the two-body photon or line final state, which we label $\langle \sigma v \rangle_{\rm line}$ (the weighting of the endpoint and continuum are then determined with respect to this cross section).
In particular, we take $\langle \sigma v \rangle_{\rm line} = \langle \sigma v \rangle_{\gamma \gamma} + \langle \sigma v \rangle_{\gamma Z}/2$, i.e. an appropriately weighted combination of the two-photon and $\gamma Z$ final states.
An extended discussion of this point can be found in Ref.~\cite{Baumgart:2017nsr}.

Under the assumptions we have adopted, at the thermal masses H.E.S.S. is sensitive to both the Wino and Quintuplet, but is more than a factor of a few away from the predicted Higgsino flux.
Assuming the Wino constitutes all of the DM mass away from the thermal value, we can probe masses up to 4 TeV.
The same logic applied to the Higgsino does allow a small mass window near the Sommerfeld peak at 6.5 TeV to be probed, and for the Quintuplet only certain mass ranges above 20 TeV are out of reach.

The lower panel of Fig.~\ref{fig:WinoHiggsinoQuintuplet} shows the sensitivity of the mean expected limit to the halo profile, but also breaks down the contribution to the limit by the various contributions to the spectrum.
As discussed in Sec.~\ref{subsec:TeVmodels}, the Higgsino spectrum consists of only the line originating from the two-photon final state, in addition to continuum emission.
For the Wino and Quintuplet, we also have accounted for the endpoint contribution.

A striking feature of the sensitivity projections is the sharp features in the expected limits placed on the Quintuplet.
For all three DM models, the theoretical cross-sections exhibit sharp features associated with Sommerfeld resonances, but only for the Quintuplet the features emerge in the limit.
As the lower panel of Fig.~\ref{fig:WinoHiggsinoQuintuplet} reveals, these features originate from sharp variations in the endpoint and continuum spectrum as a function of mass.
This point is explored further in Fig.~\ref{fig:GammaYieldWinoHiggsinoQuintuplet}, where we show spectra for three models as a function of several nearby masses.
For the Wino and Higgsino, we show masses that run across a Sommerfeld resonance, and in both cases the spectrum is seen to vary smoothly as a function of mass.
For the Quintuplet, however, when we run across the first feature seen in the expected limit we see there can be large changes in the spectrum even for very small variations in the mass.
The origin of this behavior is a competition between the various channels that the Sommerfeld process can allow the Quintuplet to annihilate through.
For the Wino, the neutral initial state $\chi^0 \chi^0$ can transition into a charged state $\chi^+ \chi^-$, which then readily annihilates to two photons.
For the Quintuplet, not only can this occur, but also the initial state can transition into a doubly charged state $\chi^{++} \chi^{--}$.\footnote{For larger SU(2) multiplets we expect that effects of this kind could be even more pronounced.
As a specific example, the Septuplet, which transforms as a $\mathbf{7}$ of SU(2), could also transition into triply charged states.
For a recent discussion of the Septuplet, see Ref.~\cite{Bottaro:2021snn}.}
Both of these final states have a spectrum of Sommerfeld resonances associated with them, and for the Quintuplet between $2-3~{\rm TeV}$ the theory transitions between annihilation being dominated by the singly and doubly charged final states; one of these processes is turning off exponentially as we move away from its associated Sommerfeld peak just as the other is turning on exponentially as we approach its peak.
This gives rise to the rapid variation in the spectrum shown.
The presence of the multiple channels also contributes to the richer structure of Sommerfeld resonances seen in the theoretical prediction of the Quintuplet cross section.

\section{Conclusions}
\label{sec:conclusions}

In this work, we have looked towards the ultimate sensitivity of the current generation of IACT instruments to annihilating DM, with a view in particular to understanding what factors will eventually limit their reach.
We have focused on a considerable range of possible DM annihilation, including a broad suite of two-body final states as well as canonical electroweak candidates, across the mass range 0.5 to 100 TeV.
The problem has been approached with state-of-the-art computations of the DM spectra as well as models for the DM density distribution in the inner Galaxy as the most privileged region of the sky for  detection of annihilating DM.
Taken together, our estimates suggest that H.E.S.S.-like instruments should be able to reach, or come within an order-of-magnitude of, the thermal relic prediction for a range of two-body channels for $m_\DM \sim 1~{\rm TeV}$.
An exception is the neutrino channels, however, we find that IACTs can provide competitive constraints on these channels to neutrino telescopes such as ANTARES.
We have further demonstrated that a H.E.S.S.-like instrument can probe the thermal Wino and Quintuplet, and have identified a rapidly varying spectrum as a function of mass as a unique feature in searches for the latter.
The thermal Higgsino, however, remains out of reach by a factor of at least a few.
In drawing these conclusions, we note that we have conservatively assumed an eventual 500 hour dataset.
We reiterate, however, that in the next two years H.E.S.S. is well positioned to reach a total of over 1,000 hours near the GC, which could push its reach even further, and suggests that CTA should be looking to collect even larger datasets.

Our analysis of the limiting uncertainties to the eventual DM reach has identified -- perhaps unsurprisingly -- that the dominant contribution is the uncertainty on $\rho_\DM$ in the inner Galaxy.
While tremendous developments in our understanding of the DM distribution in the Milky Way have occurred in the era of Gaia and ever more accurate simulations, DM searches with IACTs rely heavily on the distribution right near the dynamic center of the Galaxy, where the uncertainties are greatest.
Future improvements in our understanding of $\rho_\DM$ will immediately translate into reduced systematic uncertainties on the DM sensitivity.
Further, should the profile turn out to be more peaked near the GC, then the DM reach could be even greater than what we have projected in this work, which may have dramatic implications for the Higgsino.
Beyond the DM profile, our results demonstrate the next most important effect in terms of sensitivity is the statistical uncertainty, and this highlights the importance of continued data collection for current generation instruments.
We further explored the systematic uncertainty induced by various background sources.
The impact of a possible contribution from the Fermi Bubbles or GCE are negligible.
More important are uncertainties on the spectrum of hadronic cosmic rays and even a contribution from the GDE at these energies.
The possibility that the GDE could impact DM searches also suggests it may yet be detectable with current generation IACTs, and we leave this as an important question to be resolved by future work.

In fact our work leaves several important questions to be resolved before the ultimate reach of current generation IACTs can be determined.
For instance, in this work we have adopted an ON-OFF analysis, with the OFF region observations collected from extragalactic surveys far from the galactic plane.
But this ultimately requires significantly more observation time, and it remains important to determine whether purely Monte Carlo simulation based approaches can replace OFF observations.
An alternative approach makes uses of background models derived from Monte Carlo simulations or blank extragalactic-field observations.
Background models are being developed with current IACT data (see, for instance, Ref.~\cite{Mohrmann:2019hfq}).
This approach has been used in the context of CTA~\cite{CTA:2020qlo}. However, such an approach does not reach yet the level of control of the systematic uncertainties required to be used in the GC region~\cite{HESS:2022ygk}.  
Further, as instruments like H.E.S.S. continue to collect GC observations in the coming years, it will be important to study the optimal scan strategies to be employed, in a manner that balances both the reach for DM, but also the systematic robustness of the results.

In the present work, for our model-independent constraints, we focused on the most commonly considered scenario where the velocity-weighted DM annihilation cross section is independent of the relative velocity between the annihilating states, i.e., we have assumed the annihilation is dominantly $s$-wave.
However, in certain models, higher partial waves may dominate the annihilation process.
If so, the annihilation rate will inherit an additional dependence on the environment, becoming suppressed in regions with smaller relative velocity.
Assuming that the DM particles and stars are in equilibrium, the annihilation would be suppressed in dwarf spheroidal galaxies as compared to the Galactic Centre region, given the smaller measured stellar velocity dispersion in those systems.
The specific case of $p$-wave annihilation in the Galactic Center and the appropriate modifications one must make to the $J$-factor has been discussed in, for instance, Refs.~\cite{Boddy:2018ike,Johnson:2019hsm,Board:2021bwj,McKeown:2021sob,Kiriu:2022bjq}, and we refer to those references for further discussion.

In summary, even after many years of searches for TeV scale DM, the future remains bright.
Not only will CTA soon begin collecting its potentially revolutionary dataset, but in the near term further improvements can be expected from the current generation of IACTs.
The first hints of DM annihilation may well begin to emerge in the coming years.

\begin{acknowledgments}
We warmly thank Matthew Baumgart, Tracy Slatyer, and Varun Vaidya for sharing the gamma-ray spectra of electroweak Quintuplet DM prior to its publication~\cite{quintuplet:upcoming}.
NLR further thanks Bryan Webber for useful conversations related to \texttt{HDMSpectra}.
\end{acknowledgments}

\appendix

\begin{figure*}[!t]
\includegraphics[width=0.45\textwidth]{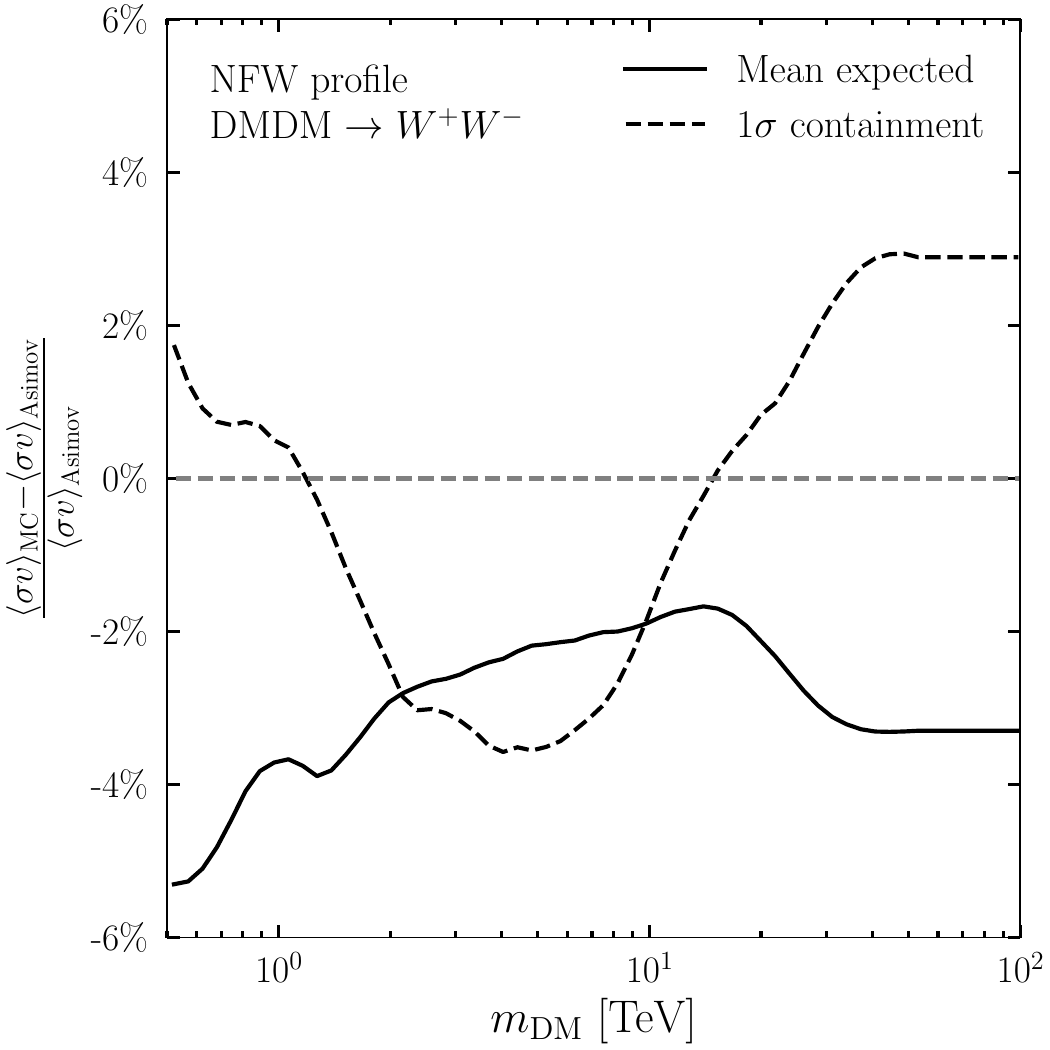}\hspace{0.5cm}
\includegraphics[width=0.45\textwidth]{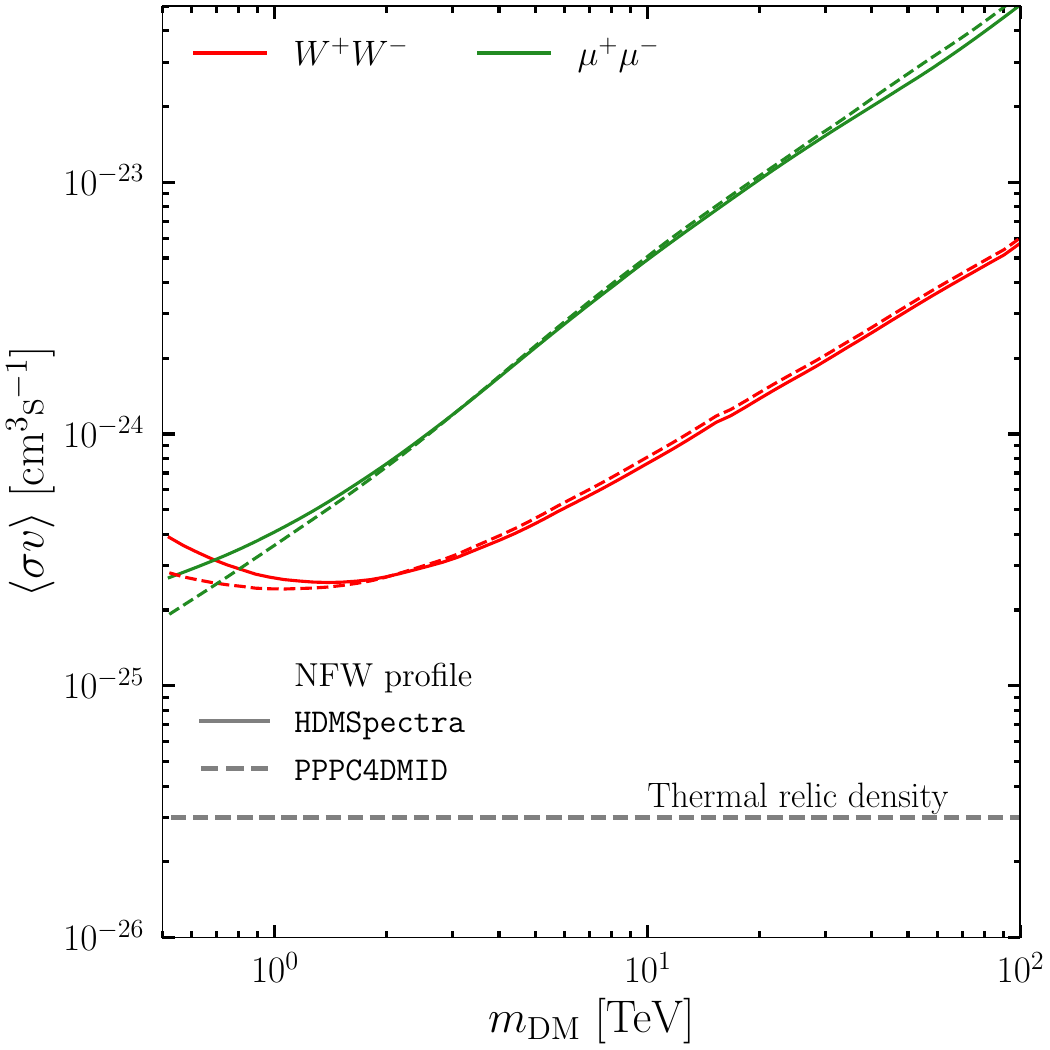}
\caption{(Left) Percentage differences between the Asimov and Monte Carlo simulation computations of the expected mean upper limits (solid line) and the 1$\sigma$ containment band (dashed line) on $\langle \sigma v \rangle$ as a function of the DM mass $m_{\DM}$.
The limits are computed at 95\% C. L. on $\langle \sigma v \rangle$ for the $W^+W^-$ channel derived for the H.E.S.S.-like mock dataset of GC observations and using the computation of the gamma-ray yield from \texttt{HDMSpectra}.
(Right) Similar to Fig.~\ref{fig:LimitsWWComparisonSpec}, but here showing results for both the $W^+ W^-$ and $\mu^+ \mu^-$ two-body final states.} 
\label{fig:ProcedureComparison}
\end{figure*}

\begin{figure*}[!t]
\includegraphics[width=0.45\textwidth]{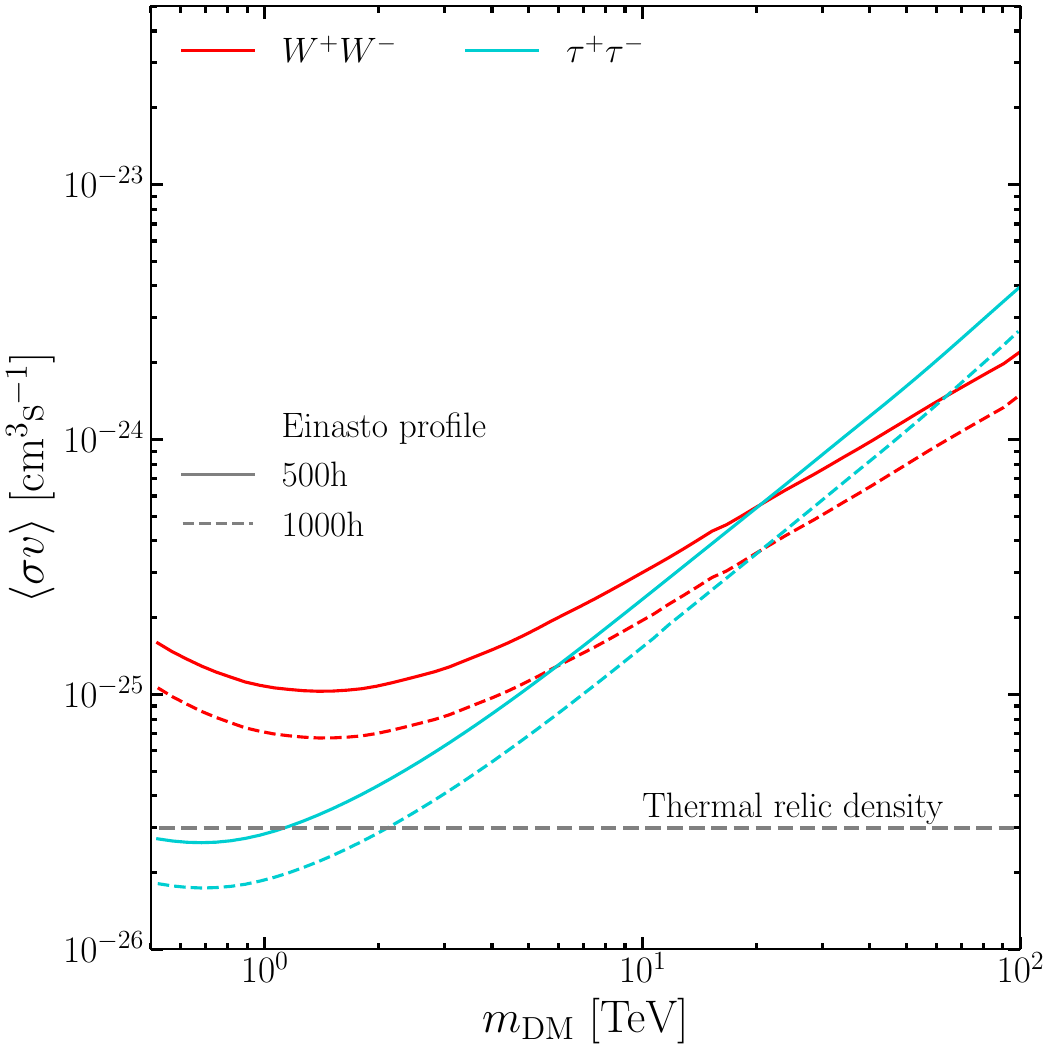}
\caption{Similar to Fig.~\ref{fig:LimitsWWComparisonSpec}, but here showing results for 500\,h and 1,000\,h of flat time exposure across the considered ON region. The sensitivity is shown for the $W^+ W^-$ and$\tau^+\tau^-$ two-body final states.} 
\label{fig:Comparison500to1000}
\end{figure*}

\section{Comparison of the Asimov procedure to a Monte Carlo based approach}
\label{sec:appendixA}

In the main body of this work, we obtained our sensitivity estimates through the use of the Asimov procedure~\cite{Cowan:2011an}.
As discussed in Sec.~\ref{subsec:Statmethod}, it is also possible to compute limits using a Monte Carlo simulation based approach.
On the left of Fig.~\ref{fig:ProcedureComparison}, we show that the Asimov approach provides an accurate determination of the sensitivity, by comparing the results it obtains to those derived from the distribution obtained across 300 simulations.
The mean expected limits and the 1$\sigma$ containment band computed with the two approaches agree within 5\% and 4\%, respectively, in the probed mass range.\\

\section{Uncertainties on the DM spectrum for different channels}
\label{sec:appendixB}

On the right of Fig.~\ref{fig:ProcedureComparison} we show the difference in the sensitivity between the two computations for the gamma-ray yield -- \texttt{HDMSpectra}~\cite{Bauer:2020jay} versus \texttt{PPPC4DMID}~\cite{Cirelli:2010xx} -- for both the $W^+W^-$ and $\mu^+\mu^-$ annihilation channels (assuming an NFW density profile).
The difference between the two computations reaches 6\% and 11\% for the $W^+W^-$ and $\mu^+\mu^-$ channels, respectively, and for a DM mass of 1 TeV, spanning from 25\% up to 5\% and from 29\% up to 1\% depending on the DM mass and for the $W^+W^-$ and $\mu^+\mu^-$ channels, respectively.
This demonstrates that the $W^+ W^-$ channel considered in the main text was representative.

\section{Prospect sensitivity with 1,000 hours}
\label{sec:appendixC}

In the main body of this work, the sensitivity limits were obtained considering a flat exposure time of 500 hours. In this section, we show results obtained when 1,000 hours are considered as distributed evenly across the inner 4$^\circ$ of the GC.
We discussed in Sec.~\ref{subsec:forecastHESS} the already existing potential time exposure on the ON region available with the H.E.S.S. instrument. Therefore, assuming 1,000 hours of observations is indeed a somewhat realistic estimate of the total amount of observation time. However, the data was accumulated during different phases of H.E.S.S. and, hence, a realistic analysis would require dedicated instrument response functions separately for phase-1 and phase-2 data. Moreover, these observations were performed towards different regions of the sky. The phase-1 dataset was collected towards a region close to the GC~\cite{Abdallah:2016ygi,Abdallah:2018qtu}. Phase-2 data were obtained with observations towards a region of the sky similar to what we are considering in this work~\cite{HESS:2022ygk}. Therefore, the currently existing data are not homogeneously spread across the considered ON region as opposed to what we assume in this work. 
In Fig.~\ref{fig:Comparison500to1000}, we compare the results obtained with 500\,h and 1,000\,h of flat time exposures across the considered ON region, considering DM particles annihilating into the $W^+W^-$ and $\tau^+\tau^-$ channels.

\bibliography{bibl}

\end{document}